\documentclass[11pt]{article}
\usepackage[a4paper,margin=20mm]{geometry}
\usepackage{graphicx}
\usepackage{fontspec}
\usepackage{amsmath,amssymb}
\usepackage{booktabs,tabularx,array,longtable}
\usepackage{enumitem,pdflscape}
\usepackage[round,authoryear]{natbib}
\makeatletter
\let\bibstyle@journalplainnat\bibstyle@plainnat
\let\OriginalParseDate\NAT@parse@date
\def\NAT@parse@date#1#2#3#4#5#6@@{%
  \def\DatePrefix{#1#2#3#4}%
  \def\UndatedPrefix{n.d.}%
  \ifx\DatePrefix\UndatedPrefix
    \def\NAT@year{n.d.}\def\NAT@exlab{{#5}}%
  \else
    \OriginalParseDate#1#2#3#4#5#6@@%
  \fi}
\makeatother
\usepackage[hidelinks]{hyperref}
\usepackage{xcolor,flafter}
\newcommand{\PRE}{\mathrm{PRE}}
\newcommand{\Ret}{\operatorname{Ret}}
\makeatletter
\newcommand{\RetP}{\@ifnextchar_{\RetP@withsub}{\operatorname{Ret}_{0.10}}}
\def\RetP@withsub_#1{\operatorname{Ret}_{0.10,#1}}
\makeatother
\newcommand{\TailGap}{\operatorname{TailGap}}

\makeatletter
\newcommand{\SectionFloatBarrier}{\ifx\@deferlist\@empty\else\clearpage\fi}
\AddToHook{cmd/section/before}{\SectionFloatBarrier}
\makeatother
\title{Pharmacy stock and post-earthquake access to medicines in Tokyo: modeling choices that change recovery judgments}
\author{\parbox{0.95\textwidth}{\centering
Juanning Liang\textsuperscript{a}, Hiroki Baba\textsuperscript{a}\\[0.5em]
\small \textsuperscript{a} Graduate School of Science and Engineering, Chuo University, 1-13-27, Kasuga, Bunkyo, Tokyo, Japan\\[0.5em]
\small Corresponding author: Juanning Liang, ljuanning@gmail.com
}}\date{}
\hypersetup{pdftitle={Pharmacy stock and post-earthquake access to medicines in Tokyo: modeling choices that change recovery judgments},pdfauthor={Juanning Liang, Hiroki Baba}}
\begin{document}\maketitle
\begin{abstract}

Community pharmacies sustain access to medicines after an earthquake only while open, staffed and stocked. Accessibility models rarely represent stock depletion, and the assumptions this requires may change conclusions about recovery. We examine these assumptions for over-the-counter (OTC) medicines supplied by 4,748 registered pharmacies in Tokyo's 23 wards during the first week after a magnitude-7.3 scenario earthquake. A two-step floating catchment area model computes walking accessibility for 2,651 grid cells (500 m) under 10,000 sampled combinations of stock duration, staffing, operation and demand. Results are expressed as Retention: post-disaster accessibility divided by the same cell's pre-disaster accessibility. First, the assumed link between stock and service, the supply a pharmacy can still provide, determines the recovery verdict. If service falls in proportion to remaining stock, mean Retention declines from 0.479 (days 2–3) to 0.273 (days 4–7). If service is maintained until stock falls below one quarter, the mean rises instead, although low-end (10th-percentile) outcomes still decline. Measuring at the start of each period rather than averaging over it also reverses the trend. Second, a simple model that rescales a fixed map by citywide supply and demand reproduces citywide patterns but cannot evaluate where pharmacies close or where capacity is added; a spatial model run once on average inputs can. Third, four floating catchment methods agree on overall rankings (rank correlation 0.81), but their lowest-ranked 10\% of cells overlap only partially (Jaccard index 0.56). Disaster assessments should report stock assumptions and low-end outcomes, and treat local priority areas as method-dependent.

Keywords: Disaster accessibility; Community pharmacies; Over-the-counter medicines; Floating catchment area; Inventory modeling; Model uncertainty

\end{abstract}

\setcounter{section}{0}
\section{Introduction}

Access to medicines is a local service that must continue when an earthquake disrupts everyday urban life. Earthquakes disrupt medicine supply while residents continue to need treatment for chronic conditions \citep{who2016emergencies}. In Suzu after the Noto earthquake, temporary pharmacy services dispensed disaster prescriptions while ordinary supply was interrupted \citep{koike2026pharmaceuticalneeds}. Registered community pharmacies also supply over-the-counter (OTC) medicines for minor injuries and common complaints. In a disaster these medicines let minor conditions be handled without medical teams \citep{shikamura2017otc}; Tokyo's disaster guideline has pharmacist teams hand them to patients who need no physician's diagnosis or treatment, which reduces the load on physicians \citep{tokyo2025disasterpharmaceuticalguideline}. Unlike prescription medicines, OTC medicines need no prescriber, so access to them depends directly on whether a nearby pharmacy is open and stocked. After the 2024 Noto Peninsula earthquake, Anamizu General Hospital and nearby pharmacies faced replenishment difficulties for a week \citep{tai2024disasterpharmacists}. After Hurricane Maria, only about one pharmacy in nine in Puerto Rico was open three days later \citep{sahota2023pharmacy}. Tokyo's pharmacy business-continuity guidance therefore treats operation and supply as linked concerns \citep{tokyohealth2014pharmacybcp}. Kita ward has also built a system for reporting pharmacy and pharmacist status in emergencies \citep{mitoya2023creationfirst}. Continuity is a spatial preparedness question: service may persist in one neighborhood and lapse in the next.

Spatial accessibility methods relate service capacity, competing demand and travel \citep{luo2003spatial,luo2009enhanced}. Disaster applications have extended them to damaged facilities and disrupted road networks \citep{ertugay2016accessibility,ceferino2025acute,kaur2026healthcare}, and pharmacy applications have added opening hours \citep{clark2025pharmacy,viana2025identifyingpharmacy}. None of these studies represents the stock that an open provider can still supply. Three choices are therefore untested when such models are used to judge recovery over the first week. How does remaining stock limit service, the supply a pharmacy can still provide: in proportion, or only once stock falls below a threshold? Is recovery measured at the start of a period or averaged over it, while supplies keep being consumed? How much spatial and distributional detail does the decision require? The answer may differ for a citywide recovery verdict, a shortlist of low-access neighborhoods and the placement of additional capacity. These choices can change the recovery judgment even when the underlying pharmacy network is held fixed.

We develop a city-scale spatial stress test of OTC medicine access through pharmacies in Tokyo's 23 special wards. It covers the first week after the official southern central Tokyo magnitude-7.3 scenario earthquake. A stress test runs a system under prescribed adverse conditions to find where and when it fails \citep{kaleta2022stress}. Here the system is the pharmacy network. The conditions span stock duration, staffing, operation and additional demand; failure is measured as lost accessibility in each neighborhood. The model is run for 10,000 combinations of these conditions. We ask three questions.

Q1: Does the judgment of late-week recovery change with how remaining stock limits service, and with whether recovery is measured at the start of a period or averaged over it?

Q2: Which results can simpler calculations reproduce, and which require spatial detail or the full set of combinations?

Q3: Do the accessibility method, the spatial unit, the length of the priority list and the sample of combinations change which locations are selected for attention or support?

The modeled service is potential access to aggregate emergency OTC medicines through the registered pharmacy (\textit{yakkyoku}) network in the national Medical Information Net. Supply, staffing, stock duration and demand follow preparedness ranges motivated by Japanese disaster-pharmacy guidance and the Noto experience. Store-sales outlets (\textit{tenpo hanbaigyo}) also sell OTC medicines but fall outside this pharmacy register, so the provider set covers only part of the OTC retail network. Demand remains at residential or daytime origins. Evacuation to shelters could concentrate medicine needs elsewhere and change the competing demand around individual pharmacies; those movements are absent from the modeled demand surface. Section 6.3 discusses how these two omissions affect interpretation.

The study makes four contributions. First, it represents stock as a state variable in effective pharmacy capacity. To our knowledge, this is the first city-scale model of post-earthquake access to pharmacy-based OTC medicines with that representation. Tokyo emergency-vehicle accessibility after earthquakes \citep{kishimoto2019accessibility} and routine pharmacy accessibility \citep{yamazaki2026healthcare} have been studied separately. Second, paired tests show how the stock-to-service rule and measurement timing change recovery judgments under shared spatial inputs. Third, model-reduction tests identify the information needed to reproduce citywide patterns, local mean responses and support-location rankings. Fourth, comparisons across accessibility methods, aggregation levels, list sizes and input samples identify sensitive priority memberships.

\setcounter{section}{1}
\section{Background}

\setcounter{subsection}{0}
\subsection{Potential accessibility and pharmacy services}

Spatial accessibility relates people to service capacity. Two neighborhoods equally close to a pharmacy face different opportunities if its capacity or competing demand differs. The two-step floating catchment area (2SFCA) method first relates each provider's supply to the population within its reach. It then sums the reachable providers' supply-to-demand ratios at each population location \citep{luo2003spatial}. The enhanced version weights nearer providers more heavily through a distance-decay function \citep{luo2009enhanced}. Pharmacy applications show that the service definition changes what the measure finds. Clark and Newing \citeyearpar{clark2025pharmacy} combined pharmacy locations with opening hours in England and Wales and traced how closures changed access. Viana et al. \citeyearpar{viana2025identifyingpharmacy} found gaps in Lisbon that shift across the day as pharmacies open and close. Zhou et al. \citeyearpar{zhou2021adaptedtwo} adapted catchments to urban–rural differences, and Roubin et al. \citeyearpar{roubin2025urbanrural} mapped spatiotemporal disparities in Vermont. These studies treat pharmacy access as a time-dependent service relationship. Stock consumption adds a further constraint. An open pharmacy may have limited medicines available, so continuity depends on how inventory enters effective capacity.

\setcounter{subsection}{1}
\subsection{Disaster disruption, inventory and time}

Disaster research has propagated damage to access in several ways. Ceferino et al. \citeyearpar{ceferino2025acute} combined building and transport disruption to evaluate hospital access in a Hayward-fault scenario for the San Francisco Bay Area. Kaur et al. \citeyearpar{kaur2026healthcare} evaluated emergency healthcare in Metro Vancouver under magnitude-9 Cascadia scenarios. Their calculation combined hospital-capacity loss with inaccessible road links caused by bridge damage. Ertugay et al. \citeyearpar{ertugay2016accessibility} modeled health-service and shelter accessibility in Thessaloniki under earthquake-related road-closure probabilities. Balomenos et al. \citeyearpar{balomenos2019coastal} and Dong et al. \citeyearpar{dong2020integrated} propagated hurricane and flood damage to road access to health services. In Tokyo, Kishimoto and Osaragi \citeyearpar{kishimoto2019accessibility} simulated emergency-vehicle access under road blockage caused by building collapse. Enderami et al. \citeyearpar{enderami2024postdisaster} distinguished proximity, availability, adequacy and acceptability of essential services after a disaster. A route may remain open while the service at its end is reduced. Kaleta et al. \citeyearpar{kaleta2022stress} used a stress test to remove physicians successively from a national care network and locate the point at which patients go untreated.

These examples examine service disruption through facility availability, staffing or transport. Inventory depletion adds a different time-dependent mechanism: a site can remain functional while its usable supply falls. For pharmacies, recovering staffing and operating conditions may therefore coexist with a declining ability to serve customers. A United States scoping review documents operating status, dispensing volumes and the slow reopening of pharmacies after hurricanes and wildfires \citep{sahota2023pharmacy}. Replenishment difficulties documented after Noto \citep{tai2024disasterpharmacists} motivate scenarios in which stock and operating capacity follow different paths. Retail inventory models examine stockholding and purchase limits under panic buying \citep{kogan2022panic}. Translating a given stock fraction into service remains a modeling choice: proportional and threshold rules imply different capacities for the same stock.

Time enters the reported outcome as well. Park et al. \citeyearpar{park2026customized} used hourly accessibility estimates and sequence analysis in New York City. They showed that collapsing time variation into one aggregate hides planning-relevant patterns. After an earthquake, accessibility may be higher at phase entry yet lower when averaged over the phase, because supplies keep being consumed.

\setcounter{subsection}{2}
\subsection{Method choice, scale and placement}

Floating catchment formulations differ in how they allocate demand and represent provider choice. Delamater \citeyearpar{delamater2013spatial} showed how facility configuration enters the estimate. He proposed the modified two-step floating catchment area method (M2SFCA) for suboptimally configured systems. Páez et al. \citeyearpar{paez2019demand} diagnosed demand and supply inflation caused by counting residents in several catchments. They proposed the balanced floating catchment area method (BFCA). Jörg and Haldimann \citeyearpar{jorg2023mhv3sfca} combined provider choice with variable catchments in the modified Huff-based variable three-step floating catchment area method (MHV3SFCA).

Related comparisons show why broad agreement can conceal local differences. Apparicio et al. \citeyearpar{apparicio2008comparing} found high overall correlations between distance measures alongside local measurement errors, including errors from spatial aggregation. Chen and Jia \citeyearpar{chen2019comparative} compared 24 2SFCA formulations. Catchment size mattered most at community scale, while distance decay became more influential at broader scales. Two further choices affect a priority list independently of the formulation. Aggregating to administrative units changes the selection unit; agreement between ward summaries does not establish agreement about individual cells. Fixing the list at $K$ locations makes membership sensitive to ties near the cutoff. Resource placement makes these distinctions concrete. For a fixed amount of additional capacity, a planner needs to know whether its location changes the evaluated gain.

\setcounter{subsection}{3}
\subsection{Uncertain inputs and model reduction}

Random inputs can describe a range of service conditions. Park and Goldberg \citeyearpar{park2023stochastic} propagated historically derived intensive-care-unit supply and mobility distributions through an enhanced 2SFCA for Greater Houston. A reliability-based floating catchment approach also treats travel time as uncertain \citep{chen2020healthcareservices}. In a preparedness study, prescribed input ranges define the conditions examined. Latin hypercube sampling (LHS) \citep{mckay1979lhs} distributes values across each range and allows paired comparisons to reuse the same conditions. Its purpose is numerical coverage of the design; empirical calibration requires observations of the underlying processes.

A related question is how much of the calculation a planner needs. A fixed spatial pattern scaled by citywide totals may reproduce a citywide map without expressing where capacity changes, and a spatial calculation with mean inputs may reproduce a mean response without its distribution. A reduction test therefore compares the response to the same intervention in each model.

The studies above leave three gaps. First, post-disaster accessibility studies represent disruption through facility damage or network failure \citep{ertugay2016accessibility,ceferino2025acute,kaur2026healthcare}, and a healthcare stress test removed physicians \citep{kaleta2022stress}. None treats medicine stock as a time-varying state within effective provider capacity. Consequently, two choices that such a representation requires have not been tested: how remaining stock translates into service, and whether recovery is measured at the start of a period or averaged over it. Second, the studies reviewed here do not examine which planning tasks a simplified accessibility calculation can support. Reproducing a citywide map does not establish that a model can evaluate pharmacy closures or capacity additions at specific sites. Third, comparisons of accessibility measures have examined overall agreement, aggregation error and parameter sensitivity under routine conditions \citep{apparicio2008comparing,chen2019comparative}. They have not examined whether the locations selected as priorities remain stable under disaster conditions, across methods, aggregation scales, list sizes and input samples. This study addresses these three gaps for pharmacy-based access to OTC medicines in Tokyo, corresponding to Q1–Q3 in Section 1.

\setcounter{section}{2}
\section{Study area and data}

The study domain covers Tokyo's 23 special wards; in the 2015 data it holds 9.5 million residents at night and 12.0 million by day. A major inland earthquake would disrupt the dense pharmacy network serving them. The Tokyo Metropolitan Government's damage assessment selects a magnitude-7.3 event beneath southern central Tokyo for its effects on central metropolitan functions \citep{tokyo2022damageestimate}. The scenario represents the near-field earthquake type that Japanese planning distinguishes from distant subduction events. The rupture lies within the Philippine Sea plate. Initiation is at about 49 km depth, with the upper edge of the strong-motion generation area at about 35 km. We use the official intensity and building-damage outputs under the assessment's winter evening condition (Fig. 1; Supplementary Section S1.2). Three damage representations describe this one event: structural damage alone, and lower and upper bounds on combined structural and fire damage. The bounds assume complete overlap and no overlap, respectively. The whole-event fire extent supplies a fixed exposure bound in every phase. PRE denotes pre-disaster accessibility: the same operator applied to pre-disaster supply and demand under the same travel setting (Section 4.1).

Previous Tokyo studies provide complementary evidence. Kishimoto and Osaragi \citeyearpar{kishimoto2019accessibility} simulated emergency-vehicle movement through roads affected by building-collapse blockage and related access to local environments. Yamazaki et al. \citeyearpar{yamazaki2026healthcare} examined proximity to pharmacies and other healthcare-related facilities across the metropolis. Mitoya et al. \citeyearpar{mitoya2023creationfirst} built a disaster resource-reporting system in Kita ward. These studies address transport, proximity and reporting; the present model addresses service continuity under stock consumption.

Spatial units and phases. The study domain contains 2,651 grid cells at 500-m resolution. Four phases follow metropolitan and ward disaster-medicine planning documents \citep{tokyohealth2012medicalreliefphases,tokyohealth2015phaseguideline,otacity2018medicalphases}. T1 covers the first six hours; T2 continues to the end of the first day. T3 spans the remainder of the first three days, and T4 continues through the first week (Supplementary Table S0B). Daytime and nighttime population surfaces are evaluated separately.

Population. Both population surfaces refer to 2015, the base year of the synthetic people-flow data used for the daytime surface, the Pseudo People Flow: Personal Attribute Dataset (Tokyo). Nighttime population comes from the 2015 census by grid cell and three age groups. The people-flow data give the location and age structure of the daytime population by grid cell, but their total falls 3.46 million short of the census daytime totals of the wards. The ward-level census totals used here do not specify a distribution across the study's grid cells. We therefore keep the people-flow surface and allocate each ward's shortfall across eligible cells, so that ward totals match the census. A ten-class ordering of cells, derived from screenshots of a Mobile Spatial Statistics population map (NTT DOCOMO) taken in July 2026, sets the allocation weights; the highest class receives 16 times the weight of the lowest (R16). The allocation preserves within-cell age ratios. Alternative ratios and baseline-proportional weights are tested in Section 5.5. Age groups reweight a common accessibility surface; age-specific journeys are not modeled. Exact totals, age bands, eligibility and fallback rules are in Supplementary Section S1.2.

Pharmacies. Locations come from the Ministry of Health, Labour and Welfare (MHLW) Medical Information Net register dated 1 December 2025. Removing records without usable coordinates leaves 4,748 pharmacy endpoints. Separately licensed pharmacies at the same address are kept. Each pharmacy carries one facility-equivalent unit of baseline supply, which scenario multipliers scale to represent relative operating capacity. The register covers licensed pharmacies only (Section 1). A later coordinate audit would add 24 records and move one; these changes are documented in Supplementary Section S1.1 and are not incorporated here.

Walking network. Travel times are computed on an OpenStreetMap (OSM) walking network distributed by Geofabrik at 80 m/min. Official road layers and the metropolitan road-blockage proxy inform segment cost penalties; affected segments receive higher travel costs and remain traversable. The reference domain contains only pharmacies and demand inside the wards; Section 5.5 tests an expanded domain with 1,210 external pharmacies and 1,826 external census cells. Table 1 summarizes the inputs and their roles.

\begingroup
\footnotesize
\setlength{\tabcolsep}{2pt}
\renewcommand{\arraystretch}{1.12}
\begin{longtable}{>{\raggedright\arraybackslash}p{\dimexpr0.96\linewidth/3-2\tabcolsep\relax}>{\raggedright\arraybackslash}p{\dimexpr0.96\linewidth/3-2\tabcolsep\relax}>{\raggedright\arraybackslash}p{\dimexpr0.96\linewidth/3-2\tabcolsep\relax}}
\multicolumn{3}{@{}p{\linewidth}@{}}{\normalsize Table 1. Inputs and their roles.}\\[6pt]
\toprule
Input & Source and vintage & Role and boundary \\
\midrule
\endfirsthead
\toprule
Input & Source and vintage & Role and boundary \\
\midrule
\endhead
Earthquake and damage & Tokyo official assessment, 2022; building-conditions survey, 2021 & Seismic intensity, building damage and fire-overlap bounds \\
Population & 2015 census; synthetic people-flow data (2015 base); screenshots of a Mobile Spatial Statistics population map, July 2026 & Residential/daytime demand; daytime ward totals match the census; the map orders the daytime additions \\
Pharmacies & MHLW Medical Information Net, 1 December 2025 & 4,748 registered pharmacy endpoints; unit baseline capacity \\
Walking network & OSM/Geofabrik network; official road/blockage layers & Travel weights at 80 m/min; soft blockage penalties \\
\bottomrule
\end{longtable}
\endgroup

\begin{figure}[!tp]\begin{minipage}{\linewidth}\centering
\includegraphics[width=170mm,height=180mm,keepaspectratio]{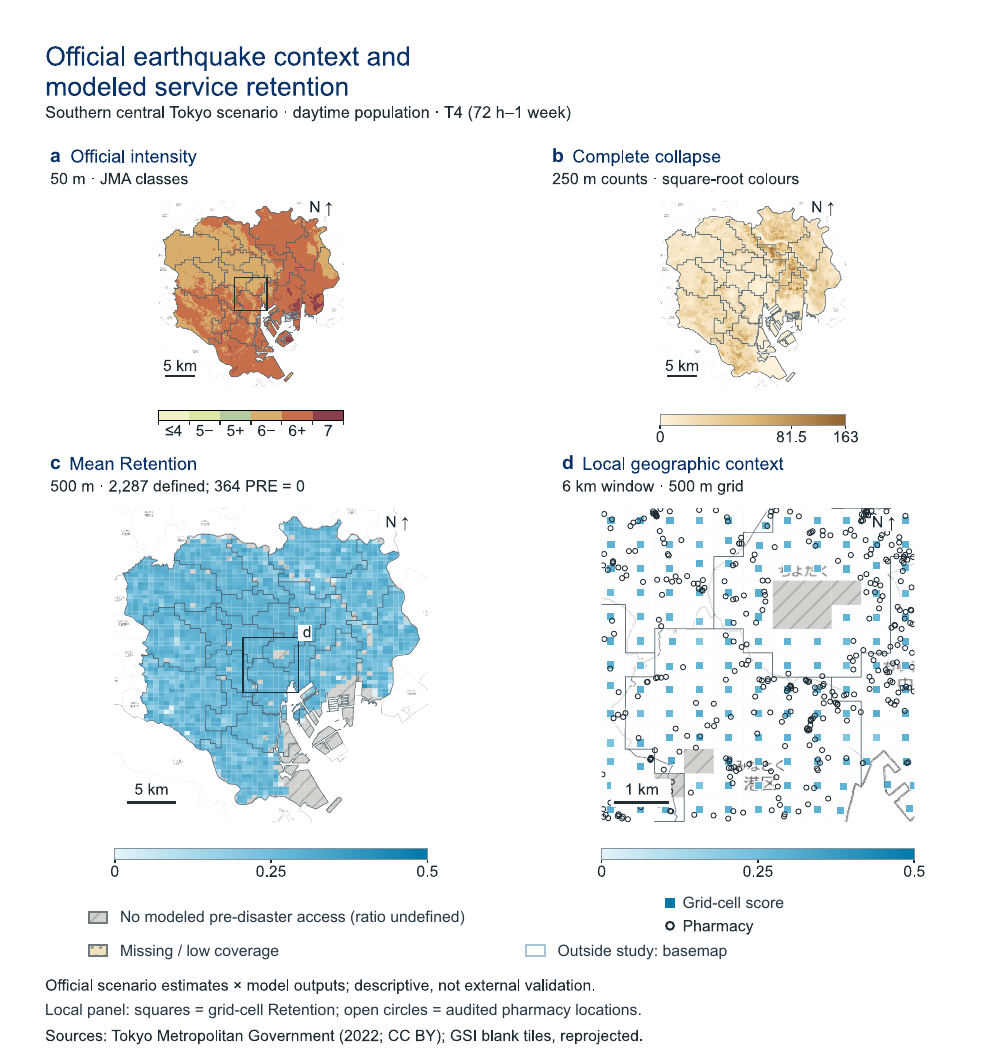}
\par\smallskip\raggedright\small Figure 1. Study area and inputs. (a) Official seismic intensity at 50-m resolution. (b) Complete-collapse counts at 250-m resolution. Both use the southern central Tokyo earthquake scenario under winter evening conditions \citep{tokyo2022damageestimate} (Supplementary Section S1.2). (c) Focal final-phase mean Retention on 2,287 positive-PRE cells; hatching marks 364 undefined cells. (d) A 6-km window of grid-cell Retention and pharmacies. JMA: Japan Meteorological Agency; GSI: Geospatial Information Authority of Japan; CC BY 4.0: Creative Commons Attribution 4.0 license. Map lines delineate study areas and do not necessarily depict accepted national boundaries.
\end{minipage}\end{figure}

\setcounter{section}{3}
\section{Methods}

The framework has five layers (Fig. 2). An input layer constructs demand, travel weights and the pharmacy set from census, damage, network and register data (Section 3). A state layer maps five sampled coordinates and pharmacy-level random streams to phase-linked supply and demand states, including stock depletion (Section 4.2). An operator layer converts these states to accessibility. The reference operator is a Gaussian two-step floating catchment area calculation; three alternatives are substituted in Section 5.4. An outcome layer expresses accessibility as Retention relative to matched PRE and summarizes draws by their mean and 10th percentile (P10). A comparison layer runs paired contrasts, reduced calculations, operator substitutions and coordinate redraws on identical inputs (Sections 4.3–4.7). The contributions lie in the state and comparison layers. Keeping a standard reference operator allows each comparison to isolate one change. Table 2 collects the terms used in the results.

\begingroup
\footnotesize
\setlength{\tabcolsep}{2pt}
\renewcommand{\arraystretch}{1.12}
\begin{longtable}{>{\raggedright\arraybackslash}p{\dimexpr0.21\linewidth-2\tabcolsep\relax}>{\raggedright\arraybackslash}p{\dimexpr0.63\linewidth-2\tabcolsep\relax}>{\raggedright\arraybackslash}p{\dimexpr0.12\linewidth-2\tabcolsep\relax}}
\multicolumn{3}{@{}p{\linewidth}@{}}{\normalsize Table 2. Terms used in the results.}\\[6pt]
\toprule
Term & Meaning & Defined in \\
\midrule
\endfirsthead
\toprule
Term & Meaning & Defined in \\
\midrule
\endhead
PRE & Pre-disaster accessibility of a grid cell, computed with the same operator and travel setting & 3, 4.1 \\
Retention & Post-disaster accessibility divided by PRE; undefined where PRE is zero & 4.1 \\
P10 & Within-cell 10th percentile over the 10,000 draws, expressed relative to PRE; a lower-tail outcome at that location, not a spatial decile & 4.1 \\
Draw & One sampled input combination, carried through all four phases & 4.2 \\
Model case & One complete calculation configuration evaluated over all draws & 4.2 \\
Profile & One of six day/night × damage-representation combinations; the focal profile uses daytime population and the lower fire-overlap bound & 4.3 \\
Full model & The framework with all draws and the Gaussian 2SFCA reference operator & 4.1 \\
Reference operator & The Gaussian enhanced 2SFCA calculation that converts supply, demand and travel weights into accessibility & 4.1 \\
Alternative operators & BFCA, M2SFCA and MHV3SFCA, defined in Section 2.3 and compared on shared inputs & 4.6 \\
\bottomrule
\end{longtable}
\endgroup

\setcounter{subsection}{0}
\subsection{Accessibility operator, effective supply and Retention}

Accessibility is calculated from three inputs: travel weights between grid cells and pharmacies, effective supply at each pharmacy, and demand in each grid cell (Fig. 2). Travel times on the walking network are converted to weights by a Gaussian decay with a 15-min cutoff. For origin grid cell i and pharmacy j in travel setting $\kappa$,

\[
W_{ij}^{(\kappa)}=\begin{cases}\exp\!\left[-\left(c_{ij}^{(\kappa)}\right)^2/\beta_\kappa\right],&c_{ij}^{(\kappa)}\le c_0,\\0,&c_{ij}^{(\kappa)}>c_0\lor\mathcal{P}_{ij}=\varnothing.\end{cases}
\tag{1}
\]

Here, $c_{ij}^{(\kappa)}$ is effective travel time and $\mathcal P_{ij}$ is the set of connecting paths. The compound index $\kappa$ identifies the road, decay and mobility specifications and population surface. The symbols $\lor$ and $\varnothing$ mean “or” and an empty path set. The dimensionless exponential weight is zero if no finite route exists or the cutoff $c_0$ is exceeded. The reference cutoff is $c_0=900$ s. The decay parameter is $\beta_\kappa=300{,}000$ $\mathrm{s}^2$. Weights are precomputed once per travel setting and reused in every draw.

Road disruption is represented for walking. Pedestrians can pass gaps that block vehicles: Kishimoto and Osaragi \citeyearpar{kishimoto2019accessibility} use clearance thresholds of 4 m for vehicles and 0 m for pedestrians. We therefore represent disruption through higher travel costs on affected road segments, using the official 250-m blockage indicator as an exposure proxy. Supplementary Section S2 gives the road classes and a hard-closure sensitivity check.

The full model combines the pharmacy-continuity stress-test framework with the reference operator and all 10,000 draws. Its implementation identifier in the Supplement is TPEP-2SFCA, where TPEP abbreviates Tokyo Post-Earthquake Pharmacy. The reference operator is the Gaussian-weighted enhanced 2SFCA \citep{luo2009enhanced}, using the same weights in both steps. For fixed phase and draw, $D_i$, $A_i$, $S_j$ and $B_j$ denote demand, accessibility, supply and catchment demand. It has two useful identities. Demand-weighted accessibility equals supply at pharmacies with positive catchment demand: $\sum_iD_iA_i=\sum_{j:B_j>0}S_j$. At fixed demand and weights, accessibility is linear in supply, so a shared stock multiplier factors out and a local capacity increment has a baseline-independent effect. Population-weighted Retention additionally depends on population and matched PRE. These properties motivate the reduction and placement tests. Comparators alter demand allocation (BFCA), configuration weighting (M2SFCA) or supply-dependent provider choice (MHV3SFCA), allowing their consequences to be tested on shared inputs. The reference calculation first divides pharmacy supply by catchment demand, then sums the weighted ratios at each grid cell:

\[
\begin{aligned}
B_{jt}^{(\ell)}&=\sum_iW_{ij}^{(\kappa)}D_{it}^{(\ell)},\\[4pt]r_{jt}^{(\ell)}&=\begin{cases}S_{jt}^{(\ell)}/B_{jt}^{(\ell)},&B_{jt}^{(\ell)}>0,\\0,&B_{jt}^{(\ell)}=0,\end{cases}\\[4pt]A_{it}^{(\ell)}&=\sum_jW_{ij}^{(\kappa)}r_{jt}^{(\ell)}.
\end{aligned}
\tag{2}
\]

Indices $t=1,\ldots,4$ and $\ell=1,\ldots,10{,}000$ identify phase and draw. Here $A_{it}^{(\ell)}$ is accessibility, $D_{it}^{(\ell)}$ is standardized demand, $S_{jt}^{(\ell)}$ is effective facility-equivalent supply, $B_{jt}^{(\ell)}$ is catchment demand and $r_{jt}^{(\ell)}$ is the supply-to-demand ratio. Providers with zero catchment demand contribute zero; origins with no weighted connection have zero accessibility.

Effective pharmacy supply combines baseline capacity, availability, staffing, operating conditions and stock:

\[
S_{jt}^{(\ell)}=S_j^0C_j^{(\ell)}F_{jt}^{(\ell)}O_{jt}^{(\ell)}\omega_t^{(\ell)}.
\tag{3}
\]

Baseline pharmacy supply is $S_j^0=1$ facility-equivalent unit. $C_j^{(\ell)}$ is one for an available pharmacy and zero otherwise. In the reference cases every pharmacy is available; ward quotas of 0.75 and 0.50 enter only the available-pharmacy comparisons (Section 4.7) and the reconfiguration test (Section 4.5). $F_{jt}^{(\ell)}$ and $O_{jt}^{(\ell)}$ are staffing/function and operating multipliers. The stock fraction $\omega_t^{(\ell)}$ ranges from one initially to zero at exhaustion and is shared by pharmacies within each draw.

In the proportional rule, service (the supply a pharmacy can still provide, as a share of its functioning capacity) is proportional to the remaining stock fraction, which depletes without replenishment:

\[
\omega_1^{(\ell)}=1,\qquad\omega_{t+1}^{(\ell)}=\max\!\left\{0,\omega_t^{(\ell)}-\frac{\Delta t_t\gamma_{\mathrm{pol}}}{H^{(\ell)}}\right\}.
\tag{4}
\]

Here $\omega_t$ is the fraction of initial stock remaining at the start of phase t. It equals one at the outset and zero at exhaustion. The depletion horizon H is the time needed to consume the initial stock at the reference rate. It is sampled between 3 and 7 days, with five days as the middle value. These bounds follow guidance anticipating medicine shortages for three to seven days after a metropolitan earthquake (Supplementary Section S5.8). Phase duration $\Delta t_t$ is measured in days: 0.25, 0.75, 2 and 4. The policy multiplier $\gamma_{\mathrm{pol}}$ slows consumption under purchase limits; its reference value is one. The recurrence reaches T4 entry at 72 h. Consumption during the following phase enters only the stock-balance phase-average extension.

Demand combines routine OTC purchases, minor-injury-related needs and needs among other affected residents. Routine demand depends on population by age group; the additional components also depend on building damage and phase-specific multipliers:

\[
D_{it}^{(\ell)}=\lambda_D\left[D_{it}^{\mathrm{base}}+D_{it}^{\mathrm{minor},(\ell)}+D_{it}^{\mathrm{affected},(\ell)}\right].
\tag{5}
\]

The common factor $\lambda_D=4748/9{,}546{,}305$ converts raw demand to the supply scale. It uses the unit-capacity pharmacy total and nighttime population. It is applied once and held fixed across all comparisons. Superscripts base, minor and affected label routine, minor-injury and affected-only (AO) demand; Supplementary Section S3 gives their coefficients.

For each grid cell and phase, the $M=10{,}000$ simulated values have mean $\bar A_{it}$ and 10th percentile $A_{it}^{0.10}$ (P10), computed by the Type-7 quantile function $Q_{0.10}$. P10 is a lower-tail (low-end) value at that location, with about 10\% of sampled outcomes at or below it; it is not the lowest-ranked 10\% of grid cells. Both are divided by matching pre-disaster accessibility $A_i^{\PRE}$, calculated with the same method and travel setting:

\[
\begin{aligned}
\bar A_{it}&=\frac1M\sum_{\ell=1}^MA_{it}^{(\ell)},&A_{it}^{0.10}&=Q_{0.10}\left(A_{it}^{(1)},\ldots,A_{it}^{(M)}\right),\\[4pt]\Ret_{it}&=\frac{\bar A_{it}}{A_i^{\PRE}},&\RetP_{it}&=\frac{A_{it}^{0.10}}{A_i^{\PRE}},\\[4pt]\TailGap_{it}&=\Ret_{it}-\RetP_{it}.
\end{aligned}
\tag{6}
\]

Here $\Ret_{it}$ and $\RetP_{it}$ are mean and P10 Retention; $\TailGap_{it}$ is their difference, reported in Fig. 4a. Retention of 0.5 means half the baseline score, not that half the residents obtained medicine. At fixed demand and travel weights, a cell has this value when every pharmacy within reach supplies half of its normal capacity. We use 0.5 as the level against which low Retention is reported. Population weighting averages grid-cell values over the eligible population. Retention is undefined where PRE is zero and is not capped at one. Priority lists rank locations by mean or P10 Retention. Descriptive tail sets keep all cutoff ties; fixed-size lists break ties by origin identifier. Supplementary Sections S4.6 and S8.4 define additional outcome and persistence measures.

\begin{figure}[!tp]\begin{minipage}{\linewidth}\centering
\includegraphics[width=170mm,height=180mm,keepaspectratio]{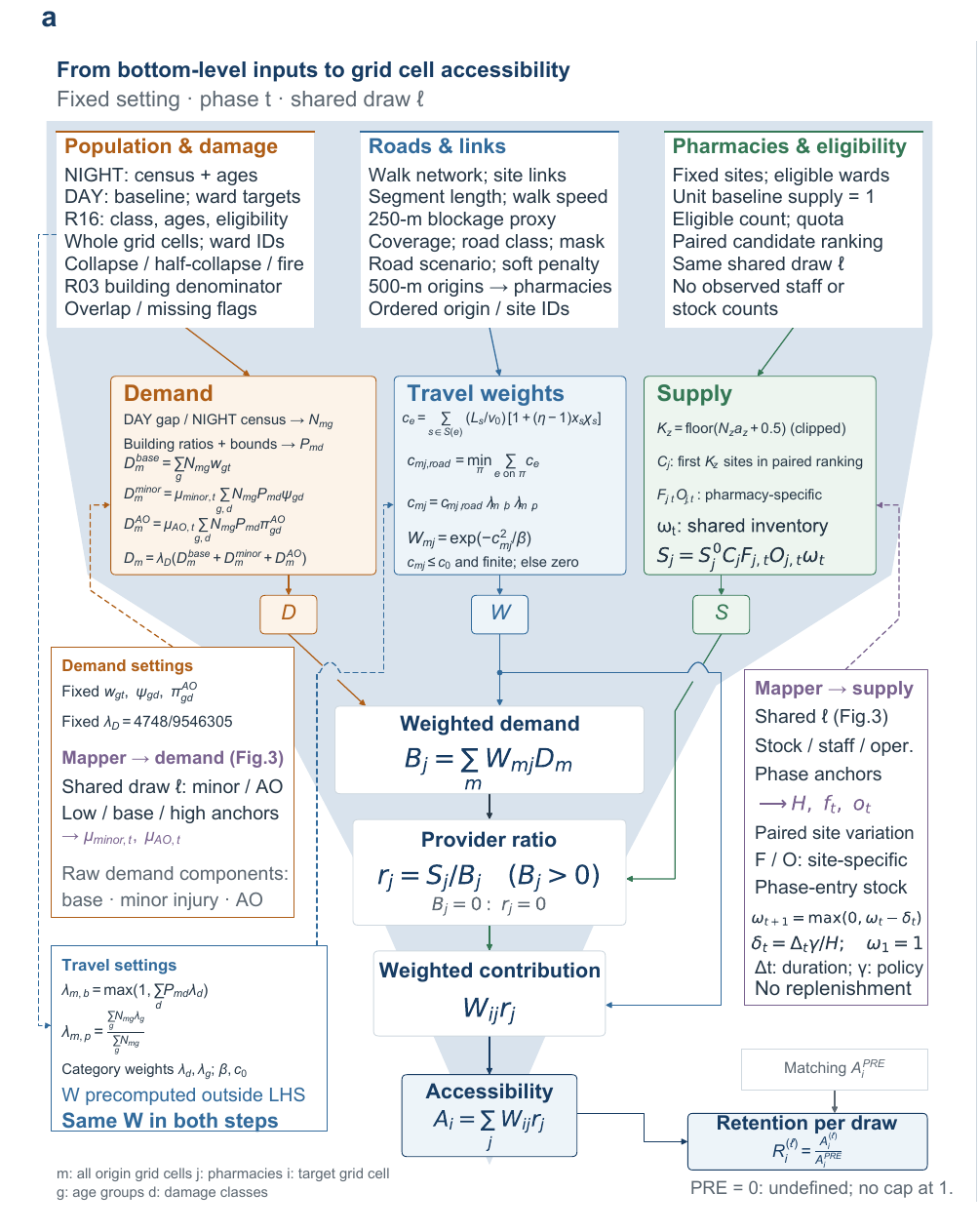}
\par\smallskip\raggedright\small Figure 2. From inputs to accessibility for one travel setting, phase and draw. Population and damage determine demand D; walking routes and blockage costs determine weights W; availability, staffing, operation and stock determine supply S. The two-step operator divides supply by weighted catchment demand, then sums accessible ratios at each origin. Retention divides accessibility by matching pre-disaster accessibility (PRE). Symbols are defined in Section 4.1 and Supplementary Table S0. The diagram covers the input, state and operator layers; outcome and comparison layers are described in Sections 4.1 and 4.3–4.7. Figure 3 explains sampled phase states. AO denotes affected-only demand; R03 denotes the 2021 (Reiwa 3) building-conditions survey.
\end{minipage}\end{figure}

\setcounter{subsection}{1}
\subsection{Sampled inputs and paired comparisons}

Staffing, operation, depletion horizons and additional demand are uncertain after an earthquake. We evaluate prescribed ranges so that comparisons cover these conditions rather than a single assumed state. Latin hypercube sampling uses one value from each of 10,000 equal-probability intervals per coordinate \citep{mckay1979lhs}. Five coordinates control the 3–7-day depletion horizon, staffing, operation, minor-injury demand and affected-only demand. They were specified before the production runs, by mechanism: stock, staff attendance and operating capacity on the supply side, and additional need per person on the demand side. The numbers of people affected follow the damage inputs and are fixed within each demand setting (Supplementary Section S3.4). Each coordinate row generates linked states across all four phases (Fig. 3; Supplementary Table S8A). Separate pharmacy-level streams add pharmacy-specific staffing and operation and, where a ward quota applies, select the available pharmacies. Paired comparisons reuse these coordinates and streams, changing only the stated assumption.

A model case is one complete calculation configuration evaluated over all draws. It specifies a phase, population surface, damage representation, travel setting, and supply and demand rules. The design comprises 660 cases spanning primary mechanisms, supply parameters and sensitivity alternatives (Supplementary Section S14). Each sensitivity alternative pairs with a reference case in the same phase, population surface and damage representation. The four operators evaluate 32 shared cases; the reference operator alone evaluates the remaining cases.

The original coordinate set was built in nested stages so that larger calculations could reuse previously computed rows (Supplementary Section S11.1). This construction preserves one-dimensional stratification but clusters joint combinations. We therefore propagated three independently generated ordinary Latin hypercube sets of 10,000 rows through the complete case design, with travel weights, input mappings and pharmacy streams held fixed. Section 5.5 reports the results.

\begin{figure}[!tp]\begin{minipage}{\linewidth}\centering
\includegraphics[width=170mm,height=180mm,keepaspectratio]{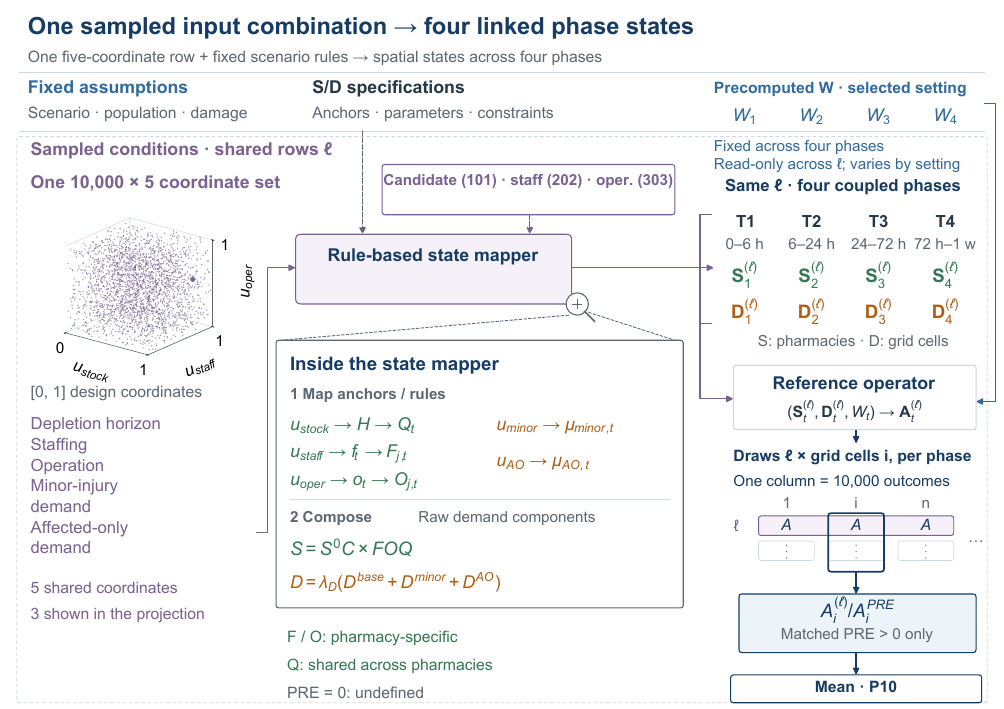}
\par\smallskip\raggedright\small Figure 3. How one sampled input combination becomes four linked phase states. Five Latin hypercube coordinates set the depletion horizon, staffing and operation trajectories, and minor-injury and affected-only (AO) demand multipliers. Fixed rules map them to linked supply and demand states across phases. Separate pharmacy-level random streams add pharmacy-specific variation and, under a ward quota, decide which pharmacies are available. Travel weights are precomputed per setting and shared by all draws. Means and P10 are computed across draws within each grid cell. Here u labels sampled coordinates, $Q_t$ is the stock fraction $\omega_t$, and n counts origin grid cells.
\end{minipage}\end{figure}

We also relate each draw’s population-weighted mean Retention to the five sampled coordinates using standardized rank regression and partial rank correlations. For the grid-cell lower tail, we compare each coordinate’s empirical distribution among bottom-decile cell–draw pairs with its uniform marginal using a Kolmogorov–Smirnov distance. These are conditional sensitivity diagnostics (Supplementary Section S5.9).

\setcounter{subsection}{2}
\subsection{Overview of the comparisons}

Table 3 maps the three research questions, and the sensitivity register of Section 5.5, to the comparisons. The citywide-surface and inventory comparisons use six profiles: daytime or nighttime population crossed with the three damage representations. The focal profile reported in figures is daytime population with the lower fire-overlap bound.

\begingroup
\footnotesize
\setlength{\tabcolsep}{2pt}
\renewcommand{\arraystretch}{1.12}
\begin{longtable}{>{\raggedright\arraybackslash}p{\dimexpr0.96\linewidth/4-2\tabcolsep\relax}>{\raggedright\arraybackslash}p{\dimexpr0.96\linewidth/4-2\tabcolsep\relax}>{\raggedright\arraybackslash}p{\dimexpr0.96\linewidth/4-2\tabcolsep\relax}>{\raggedright\arraybackslash}p{\dimexpr0.96\linewidth/4-2\tabcolsep\relax}}
\multicolumn{4}{@{}p{\linewidth}@{}}{\normalsize Table 3. Comparisons by research question.}\\[6pt]
\toprule
Question & What is compared & What is held fixed & Evaluated set \\
\midrule
\endfirsthead
\toprule
Question & What is compared & What is held fixed & Evaluated set \\
\midrule
\endhead
Q1 Does the stock-to-service rule change the recovery pattern? & Proportional versus threshold rule; longer horizon; entry top-up & Paired inputs; proportional-rule demand convention & Six profiles, T3 and T4 \\
Q1 Does the measurement timing change the apparent recovery? & Capacity at phase entry versus mean over the phase; replenishment and delayed delivery & One stock-balance process; daily-rate demand convention & Six profiles, T3 and T4 \\
Q2 How much model detail is needed? & Full model versus common-time-factor, paired scalar, mean-input spatial and deterministic calculations & Same task, inputs and support & Citywide surfaces (six profiles × four phases); T4 reconfiguration; two placement proposals \\
Q3 Do operators, scale and list size change the selected locations? & Reference operator, BFCA, M2SFCA, MHV3SFCA; grid versus ward; lists of 25, 50, 100 cells & Same case and input states; each operator's own PRE & 32 shared cases \\
Q3 Do the findings depend on the sampling arrangement? & Original nested coordinate set versus three ordinary LHS sets & Pharmacy streams, mappings, references & 660 cases; 504 contrasts; 32 shared cases \\
Sensitivity (Section 5.5): which assumptions move the map, and does the boundary matter? & 504 alternative–reference pairs in eight families; closed versus expanded domain & Matched eligible cells; recomputed PRE & 504 contrasts; six profiles \\
\bottomrule
\end{longtable}
\endgroup

\setcounter{subsection}{3}
\subsection{Inventory rules and measurement timings (Q1)}

The proportional rule makes service proportional to the remaining stock fraction $\omega_t$. A threshold alternative uses $\min(\omega_t/c_{\mathrm{plat}},1)$. Capacity remains fully available while stock exceeds $c_{\mathrm{plat}}$ of its initial level and declines proportionally below it. The main text uses $c_{\mathrm{plat}}=0.25$; Supplementary Section S5 reports alternative thresholds. With half the initial stock, the proportional rule gives half capacity and the threshold rule gives full capacity. Within this proportional-rule convention, we evaluate a five-day horizon extension and an entry top-up of 40\% of initial stock.

The measurement-timing comparison uses a separate stock-balance calculation. It allocates demand to pharmacies with row-normalized travel weights, then tracks consumption within each phase subject to stock, demand and functional capacity. Capacity is reported both at phase entry and as an average over the phase. This calculation interprets proportional-rule demand as a normalized daily rate. Thus switching from the proportional-rule convention to the stock-balance rows also changes the demand-time convention. Comparisons within each convention are read separately. Within the stock-balance process, we compare no replenishment with a 20\% addition of initial stock on day 3. A further case assigns unequal stock and delays delivery to the more exposed half of pharmacies in each ward until day 5. These comparisons isolate the service rule, measurement timing and replenishment assumptions. Equations are in Supplementary Sections S5 and S13.

\setcounter{subsection}{4}
\subsection{Reduced calculations, reconfiguration and placement (Q2)}

A reduced calculation is useful only if it preserves the response needed for the planning task. Table 4 defines the alternatives. We test their ability to reproduce citywide mean and P10 surfaces, local responses to pharmacy reconfiguration, and gains from equal-capacity placements. Reconfiguration replaces random pharmacy loss with hazard-concentrated loss at the same rounded 75\% ward quotas. Placement adds 0.25 effective-capacity units at each of 25 existing pharmacies, selected by population reach or exposure-weighted reach. Each model's response is the difference between its own endpoints. We compare population-weighted mean absolute error (MAE) and membership changes in 229-cell lists.

At fixed demand and weights, adding capacity $\Delta S_j$ at pharmacy j changes accessibility by $\sum_jW_{ij}\Delta S_j/B_j$, independently of baseline supply. The exact marginal gain in population-weighted mean Retention is $g_j=N^{-1}\sum_{i:A_i^{\PRE}>0}N_iW_{ij}\,\operatorname{E}[1/B_j]/A_i^{\PRE}$, where $N_i$ is grid-cell population, $N$ is the eligible total, and $\operatorname{E}$ averages over draws, with zero contribution where $B_j=0$. Selecting the 25 largest gains gives the exact optimum under the stated placement constraint. We compare the two reach-based proposals with this optimum using absolute and relative regret (Supplementary Section S7.6). Mean-input denominators differ from the draw-specific denominators, so their agreement with the full calculation is tested numerically.

\begingroup
\footnotesize
\setlength{\tabcolsep}{2pt}
\renewcommand{\arraystretch}{1.12}
\begin{longtable}{>{\raggedright\arraybackslash}p{\dimexpr0.96\linewidth/4-2\tabcolsep\relax}>{\raggedright\arraybackslash}p{\dimexpr0.96\linewidth/4-2\tabcolsep\relax}>{\raggedright\arraybackslash}p{\dimexpr0.96\linewidth/4-2\tabcolsep\relax}>{\raggedright\arraybackslash}p{\dimexpr0.96\linewidth/4-2\tabcolsep\relax}}
\multicolumn{4}{@{}p{\linewidth}@{}}{\normalsize Table 4. The calculations compared in Section 5.3.}\\[6pt]
\toprule
Calculation & What it does, in plain terms & Can it show where capacity has moved? & Does it give a spread of outcomes (P10)? \\
\midrule
\endfirsthead
\toprule
Calculation & What it does, in plain terms & Can it show where capacity has moved? & Does it give a spread of outcomes (P10)? \\
\midrule
\endhead
Full model & Recalculates accessibility from the supply and demand of every pharmacy and grid cell, for each of the 10,000 draws & Yes & Yes \\
Common-time-factor model & Takes a fixed T1 map and scales it by the citywide ratio of total supply to total demand in each draw & No & Yes \\
Paired scalar model & The same fixed map, scaled by each configuration's own totals & No & Yes \\
Mean-input spatial model & Recalculates accessibility once, using each location's mean supply and mean demand over the draws & Yes & No \\
Deterministic middle-input model & One calculation at the middle value of every input, without pharmacy-level randomness & Yes, for that one input case & No \\
\bottomrule
\end{longtable}
\endgroup

Note. The two fixed-map calculations keep the draw-to-draw variation of their scaling factor, which is why they still produce a P10.

\setcounter{subsection}{5}
\subsection{Operators, scale, list size and coordinate redraws (Q3)}

Three alternative operators are evaluated alongside the reference operator on the shared cases, each with its own formulation and matching PRE. These are BFCA \citep{paez2019demand}, M2SFCA \citep{delamater2013spatial} and MHV3SFCA \citep{jorg2023mhv3sfca}, introduced in Section 2.3; Supplementary Section S4 gives their settings. Pairwise Spearman correlation measures rank agreement; Jaccard overlap compares membership of bottom-decile and smaller tail sets. Binary membership entropy maps local disagreement, and ward aggregation provides the scale comparison. We also test fixed-size lists of $K$ locations (Section 5.4). Supplementary Section S8.5 cross-evaluates operator-derived support proposals. Coordinate redraws are assessed through aggregate Retention, grid-cell P10, phase directions, list membership and placement preferences.

\setcounter{subsection}{6}
\subsection{Specification families and study boundary}

The 504 paired contrasts replace one assumption at a time in eight families: road-cost rule, distance decay, additional demand, building- and population-based travel multipliers, available-pharmacy fraction, stock-use policy and building-damage denominator. Each contrast is summarized by its magnitude, the spatial 95th percentile of the absolute Retention difference, and its coverage, the share of eligible cells whose absolute difference reaches 0.05 (Supplementary Section S6). Additional comparisons vary daytime allocation, baseline capacity at fixed total, and concentrated versus random pharmacy loss. One further check lowers the ward quota to 0.25 and 0.15. Another keeps only the pharmacies that the register lists as open at three clock times. A third shortens the depletion horizon in proportion to the share of pharmacies available, as if those pharmacies absorbed all the demand of the others. Boundary sensitivity compares closed, supply-only, competition-only and full domains, with PRE recomputed in each (Supplementary Section S9).

\setcounter{section}{4}
\section{Results}

Section 5.1 describes the focal profile (Figs. 4–6). Section 5.2 answers Q1 (Fig. 7; Tables 5 and 7), Section 5.3 answers Q2 (Figs. 8–9) and Section 5.4 answers Q3 (Figs. 10–11; Table 6). Section 5.5 reports the robustness tests (Fig. 12). Six-decimal values are in the Supplement.

\setcounter{subsection}{0}
\subsection{Focal first-week profile}

The focal profile shows an early recovery followed by a late-week decline. It uses daytime population, the lower combined damage bound, proportional service and no replenishment. Population-weighted mean Retention rises through the first three phases to 0.479 at T3, then falls to 0.273 at T4. All eligible daytime residents live in cells with mean Retention below 0.5 at T1 and T4, and 54.4\% do at T3 (Supplementary Table S1). The population-weighted average of grid-cell P10 falls from 0.396 to 0.088 (Fig. 4). Both measures follow this pattern in 2,261 cells with positive PRE (Supplementary Table S8C).

Component substitutions locate the source of the decline. Restoring pre-decline inventory raises final-phase mean Retention by 0.317 and the population-weighted average of grid-cell P10 by 0.429 (Fig. 4d). Reverting staffing and operation lowers mean Retention by 0.048. Reverting demand lowers it by 0.004; the available-pharmacy set and travel weights stay unchanged. These effects are non-additive (Fig. 4d) because the operator is a ratio and supply states combine multiplicatively. Inventory drives the decline within the prescribed input ranges. Low but positive stock dominates the lower tail. Among final-phase bottom-decile cell–draw pairs, 76\% have at most one-tenth of initial stock remaining. Only 8\% of all draws meet that condition (Fig. 4c).

The phase trajectories show how operating recovery coexists with falling effective supply (Fig. 5). Between the last two phases, median stock fraction halves and effective supply falls by 39\%. The combined staffing–operation index rises by 21\%. These quantities change together under one profile. The substitution test isolates their separate effects, and Supplementary Section S5.2 records the phase values.

The 23 wards differ in ways that enter the model (Fig. 1). The five central wards (Chiyoda, Chuo, Minato, Shinjuku and Shibuya) hold 30.9\% of the modeled daytime population but 10.4\% of nighttime residents. Mean intensity and complete-collapse counts per covered cell are highest in the eastern and southern groups. At T4, population-weighted mean Retention is 0.284 in the central wards, 0.262 in the east, 0.264 in the south, 0.276 in the west and 0.278 in the north (Supplementary Tables S29–S30). Retention is lowest in the east and highest in the central wards, where the T3-to-T4 decline is also largest, 0.218 against 0.193 in the east (Fig. 6; Supplementary Table S29). The decline occurs in every profile and every operator in the matched cross-operator comparison (Supplementary Fig. S9; Table S8C).

\begin{figure}[!tp]\begin{minipage}{\linewidth}\centering
\includegraphics[width=170mm,height=180mm,keepaspectratio]{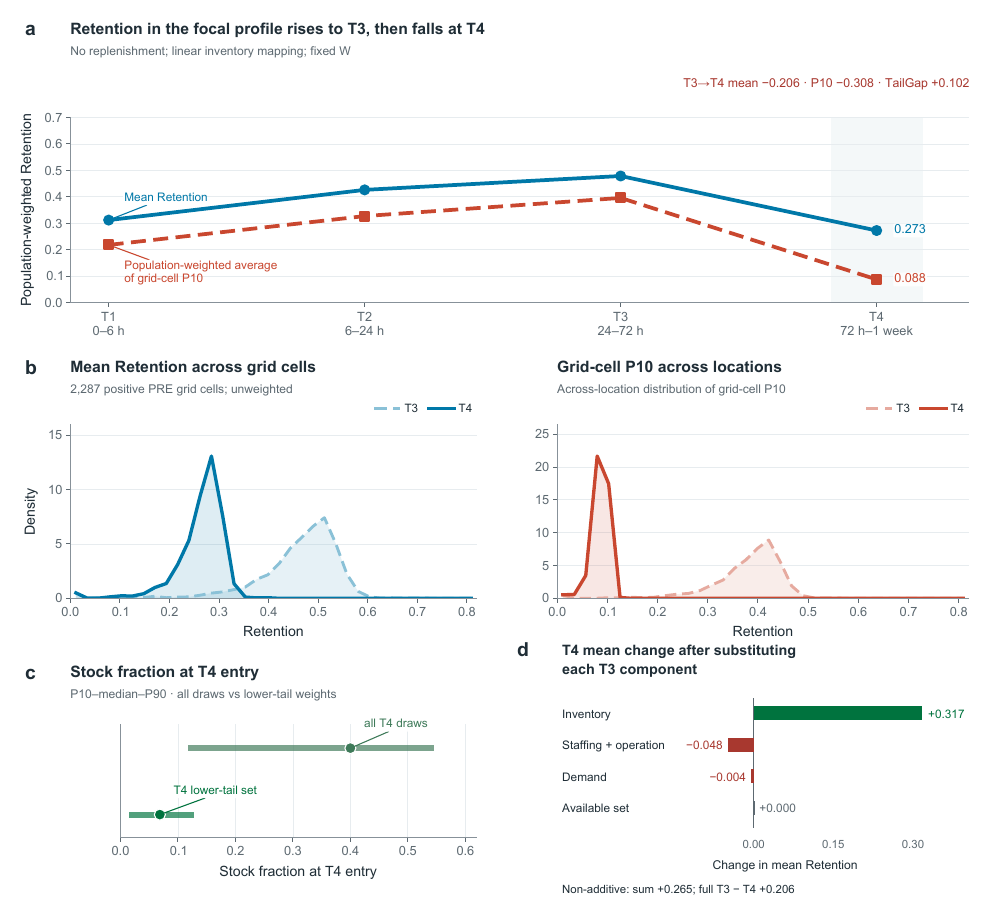}
\par\smallskip\raggedright\small Figure 4. Focal first-week profile: focal daytime population, proportional stock rule and no replenishment. (a) Population-weighted mean Retention and population-weighted average of grid-cell P10. (b) Unweighted distributions of grid-cell means and P10 at T3 and T4 over 2,287 cells with positive PRE. (c) T4-entry stock fraction, counting draws once or weighting by lower-tail cell–draw membership. (d) Change in T4 mean Retention after substituting each T3 component separately. These non-additive effects sum to +0.265; restoring the full T3 state raises Retention by +0.206. P90 denotes the 90th percentile.
\end{minipage}\end{figure}

\begin{figure}[!tp]\begin{minipage}{\linewidth}\centering
\includegraphics[width=170mm,height=180mm,keepaspectratio]{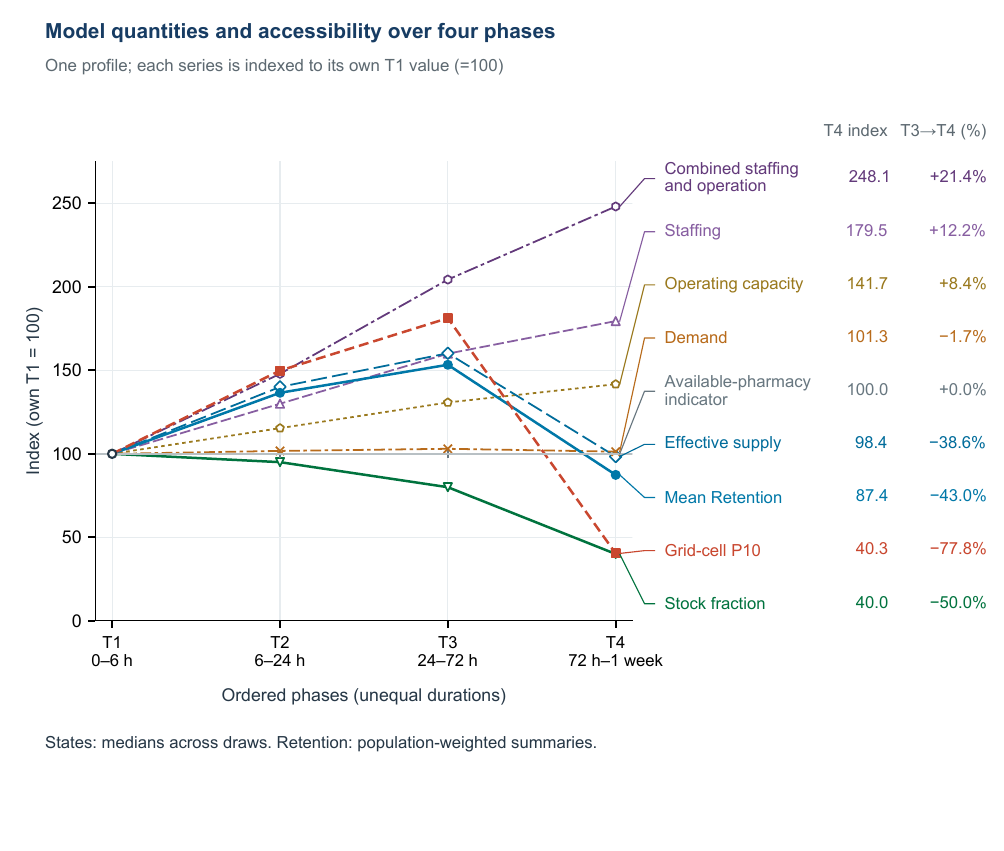}
\par\smallskip\raggedright\small Figure 5. What moved between phases in the focal profile. Seven model quantities use medians across draws; two Retention summaries use population weights. Each is indexed to its own first-phase value, so lines show changes within one scenario. Combined staffing and operation is the pharmacy-level product, not the product of the displayed lines. The right-hand columns give the final-phase index and the change from the preceding phase. T1 = 100 is an index base, not pre-disaster accessibility.
\end{minipage}\end{figure}

\begin{figure}[!tp]\begin{minipage}{\linewidth}\centering
\includegraphics[width=170mm,height=180mm,keepaspectratio]{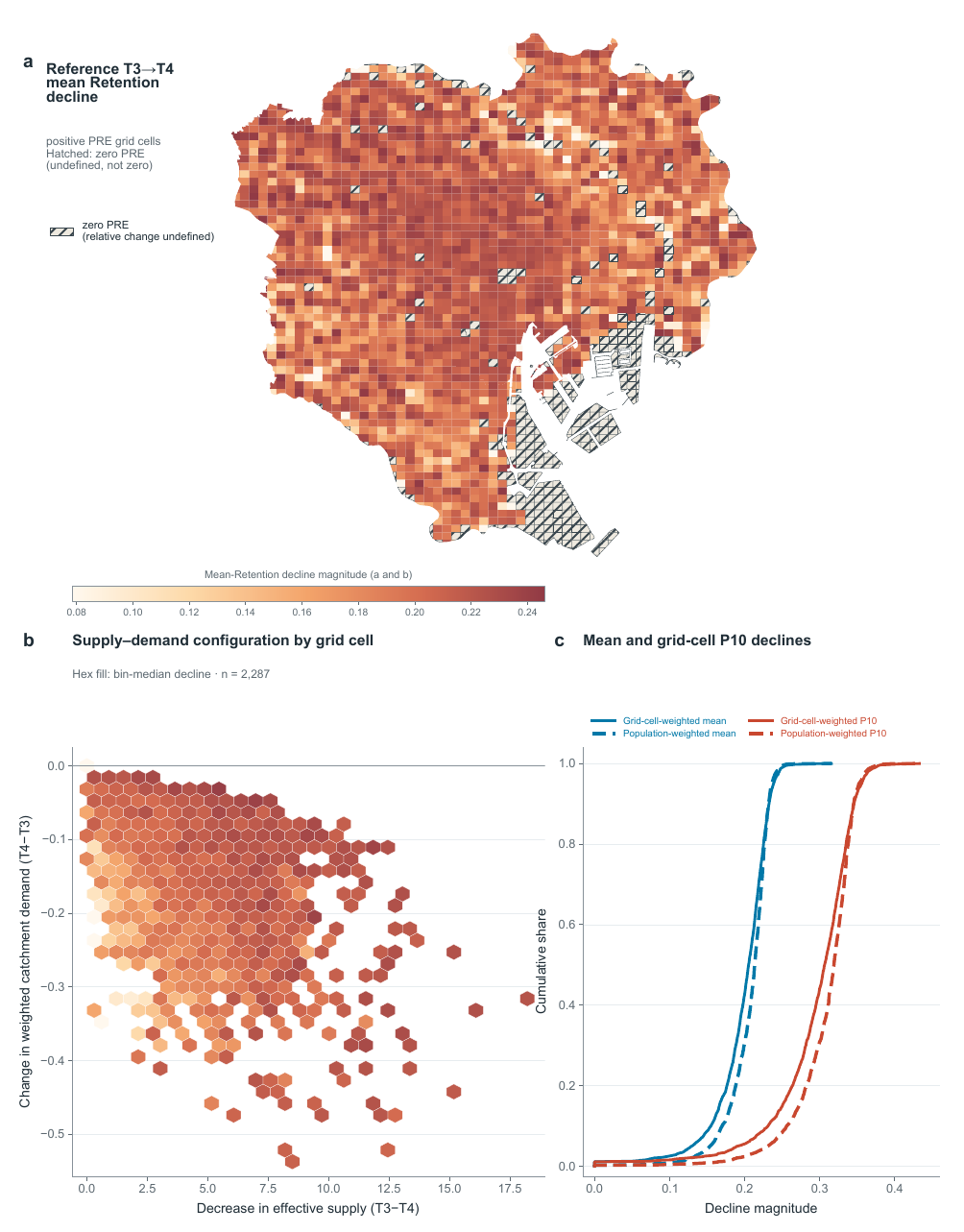}
\par\smallskip\raggedright\small Figure 6. Where the focal-profile T3-to-T4 decline occurs (focal daytime profile, 2,287 cells with positive PRE). (a) Map of the decline in mean Retention; hatching marks cells with undefined Retention. (b) Decline against the decrease in effective supply within reach (horizontal) and the change in travel-weighted competing demand (vertical); hexagon color is the median decline of the cells in each bin. The decline grows with supply loss and with rising competition, and the map shows a broad spatial gradient. (c) Distributions of the mean and P10 declines with grid-cell and population weights.
\end{minipage}\end{figure}

\setcounter{subsection}{1}
\subsection{The stock-to-service rule and the measurement timing decide the recovery verdict}

Replacing the proportional rule with a threshold rule reverses the mean recovery trend. Under the threshold rule, capacity stays fully available while stock exceeds one-quarter of its initial level. The late-week population-weighted mean therefore rises from 0.608 to 0.664 (Fig. 7). Most draws still hold stock above the threshold at the final phase's entry, while staffing and operation keep improving. The population-weighted average of grid-cell P10 nevertheless falls from 0.512 to 0.353. Draws with short depletion horizons exhaust their stock under either rule. A mean improvement therefore does not imply a lower-tail improvement under the same specification.

The measurement timing changes the recovery judgment within a single stock-balance process. Phase-entry mean Retention rises from 0.608 to 0.748. The phase-average mean instead falls to 0.569 from the same starting value. Its lower-tail average also declines (Table 5). Replenishment improves both final-phase summaries, and unequal stock with delayed delivery gives smaller gains. Within the proportional-rule convention, extending the depletion horizon and adding stock at phase entry also improve final-phase mean and P10 Retention (Table 5). In the scalar family, the entry addition needed to avoid a late-week mean decline decreases smoothly as the depletion horizon increases. This boundary is almost identical across the profiles (Fig. 7c).

\begingroup
\scriptsize
\setlength{\tabcolsep}{2pt}
\renewcommand{\arraystretch}{1.12}
\begin{longtable}{>{\raggedright\arraybackslash}p{\dimexpr0.96\linewidth/5-2\tabcolsep\relax}>{\raggedright\arraybackslash}p{\dimexpr0.96\linewidth/5-2\tabcolsep\relax}>{\raggedright\arraybackslash}p{\dimexpr0.96\linewidth/5-2\tabcolsep\relax}>{\raggedright\arraybackslash}p{\dimexpr0.96\linewidth/5-2\tabcolsep\relax}>{\raggedright\arraybackslash}p{\dimexpr0.96\linewidth/5-2\tabcolsep\relax}}
\multicolumn{5}{@{}p{\linewidth}@{}}{\normalsize Table 5. T3 and T4 summaries under alternative inventory rules and measurement timings (focal daytime profile; population-weighted; P10 average = population-weighted average of grid-cell P10).}\\[6pt]
\toprule
Specification & Demand-time convention & T3 mean & T4 mean & T4 P10 average \\
\midrule
\endfirsthead
\toprule
Specification & Demand-time convention & T3 mean & T4 mean & T4 P10 average \\
\midrule
\endhead
Proportional rule & Proportional & 0.479 & 0.273 & 0.088 \\
Proportional, depletion horizon +5 days & Proportional & 0.546 & 0.521 & 0.464 \\
Threshold rule, $c_{\mathrm{plat}} = 0.25$ & Proportional & 0.608 & 0.664 & 0.353 \\
Proportional, entry top-up 40\% & Proportional & 0.479 & 0.573 & 0.388 \\
Stock balance, phase entry & Daily rate & 0.608 & 0.748 & 0.681 \\
Stock balance, phase average & Daily rate & 0.608 & 0.569 & 0.288 \\
Stock balance, +20\% on day 3, phase average & Daily rate & 0.608 & 0.659 & 0.447 \\
Stock balance, unequal stock and delayed delivery, phase average & Daily rate & 0.604 & 0.617 & 0.445 \\
\bottomrule
\end{longtable}
\endgroup

Note. Rows with the proportional-rule convention share the proportional-rule demand; stock-balance rows share one daily-rate process. Six-decimal values are in Supplementary Section S5.6.

\begin{figure}[!tp]\begin{minipage}{\linewidth}\centering
\includegraphics[width=170mm,height=180mm,keepaspectratio]{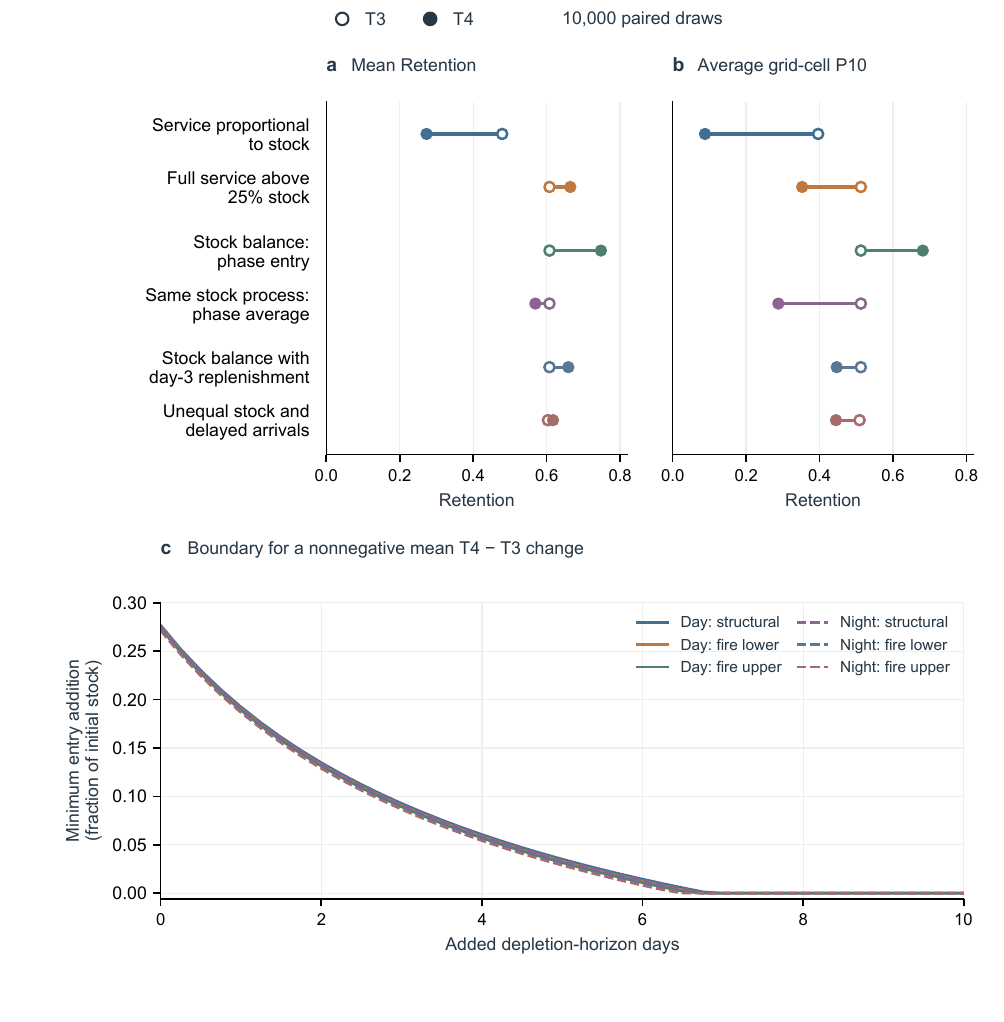}
\par\smallskip\raggedright\small Figure 7. Inventory rules and measurement timings in the focal daytime profile; open and filled symbols show the earlier and final phases. Panels show (a) population-weighted mean Retention and (b) population-weighted average of grid-cell P10. Rows compare service rules under the proportional-rule convention, then measurement timings and replenishment within one stock-balance process (Table 5). (c) Minimum entry addition needed to avoid a late-week mean decline as a function of added depletion-horizon days. Curves represent the six day/night–damage profiles.
\end{minipage}\end{figure}

Rank sensitivity shifts from staffing and operation to the depletion horizon over the week. Staffing has a squared standardized coefficient of 0.706 in the first phase; the squared depletion-horizon coefficient reaches 0.968 in the final phase (Table 7). Lower-tail outcomes also concentrate at short depletion horizons (Supplementary Fig. S12).

\setcounter{subsection}{2}
\subsection{How much model is needed: citywide surfaces, reconfiguration and placement}

Citywide surfaces are closely reproduced by a fixed-pattern calculation. The common-time-factor model has population-weighted mean-surface error no greater than 0.0085 Retention units. Its P10 error is at most 0.0077, and its bottom-decile overlap with the full model is 0.970 for the mean and 0.957 for P10. The deterministic middle-input calculation is less accurate, with final-phase error up to 0.038. Supplementary Tables S13–S14 give every profile and phase.

Local reconfiguration needs spatial inputs. Replacing random with hazard-concentrated pharmacy loss at unchanged ward quotas changes which providers remain available. The full model replaces 76 members of the mean-ranked priority list and 103 of the P10-ranked list. The paired scalar model replaces none because it changes only totals on a fixed map. Its population-weighted response error remains about 0.03 Retention units across profiles. The mean-input spatial model reproduces every mean-list replacement, with response error below $10^{-5}$ (Fig. 8). Sampling provides the corresponding P10-list changes.

Support placement also needs spatial inputs. Population-reach placement gives a larger mean-Retention gain than exposure-weighted reach in every profile (Fig. 9). In the focal profile, the gains are 0.00286 and 0.00105, respectively. The mean-input spatial model reproduces these gains at the reported precision. The scalar model assigns both proposals the same gain, 0.00122, because it cannot register where capacity is added. Supplementary Table S12 gives all profiles.

Direct optimization improves on both reach-based proposals. Ranking pharmacies by exact marginal gain gives a focal mean-Retention gain of 0.00632 under the same budget and fixed demand. The better reach-based proposal obtains only 45\% of this attainable gain. Supplementary Table S20 reports both proposals' regret. Across every profile, the optimal set shares no pharmacy with either reach-based set. The reason is that the mean objective weights an increment by inverse PRE and catchment demand as well as population.

\begin{figure}[!tp]\begin{minipage}{\linewidth}\centering
\includegraphics[width=170mm,height=180mm,keepaspectratio]{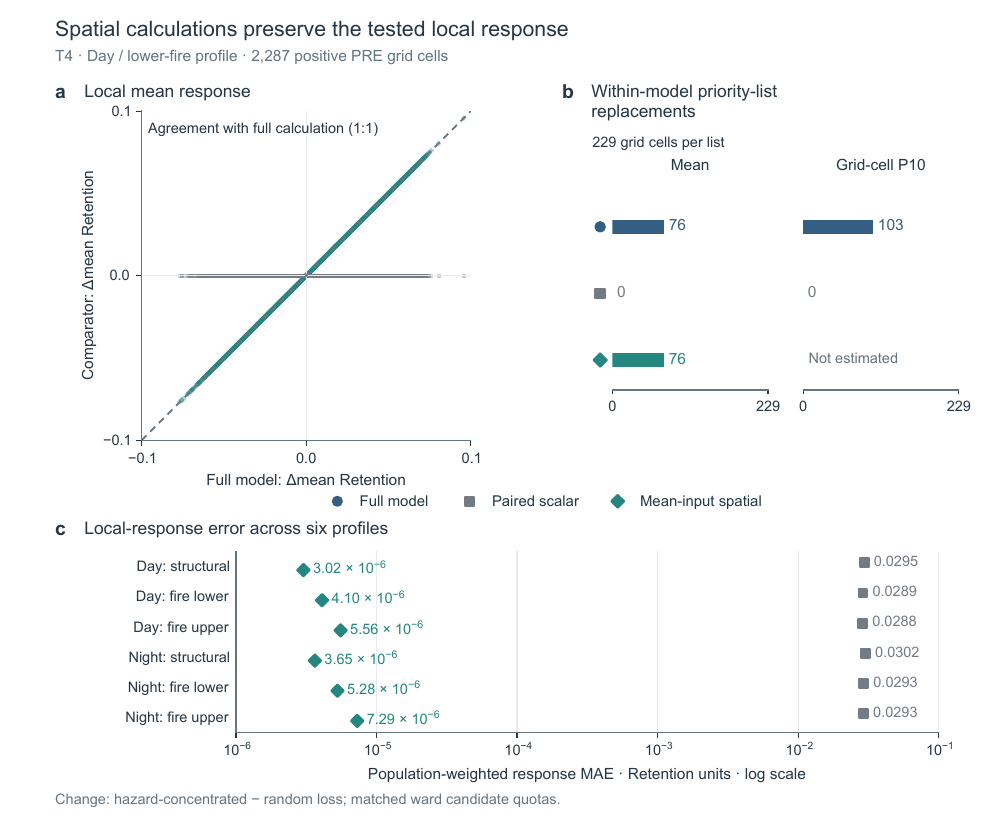}
\par\smallskip\raggedright\small Figure 8. Spatial calculations preserve the tested local response (T4, focal daytime profile). (a) Change in mean Retention after replacing random with hazard-concentrated pharmacy loss, for the paired scalar and mean-input spatial models against the full model (1:1 line). (b) Members replaced in each model's 229-cell priority lists ranked by mean and by P10; the mean-input model has no P10. (c) Population-weighted response error across six profiles (log scale).
\end{minipage}\end{figure}

\begin{figure}[!tp]\begin{minipage}{\linewidth}\centering
\includegraphics[width=170mm,height=180mm,keepaspectratio]{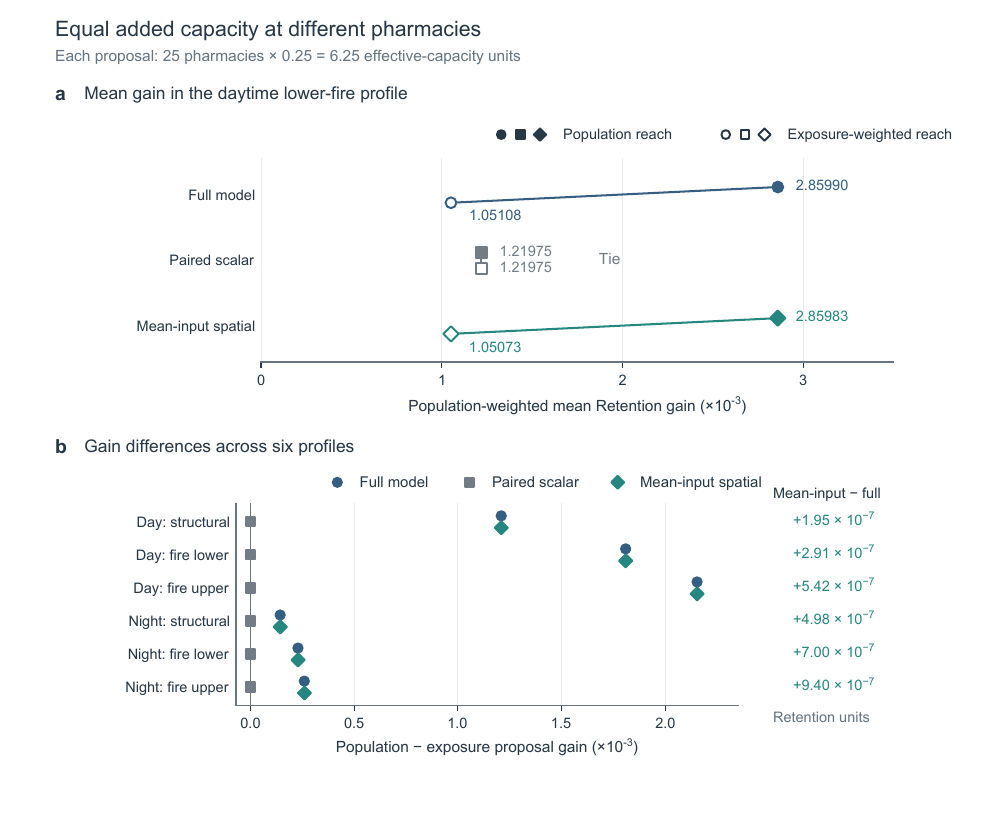}
\par\smallskip\raggedright\small Figure 9. Equal added capacity at existing pharmacies in the final phase. Each reach-based proposal adds 0.25 effective-capacity units at 25 pharmacies. The total addition is 6.25 units. (a) Focal mean-Retention gains under full, paired scalar and mean-input spatial calculations; the scalar ties the proposals. (b) Population-reach minus exposure-reach gains across profiles; mean-input and full-model gains differ by less than $10^{-6}$. Gains are additive and independent of initial supply at fixed weights and demand (Supplementary Section S7.5). Table S20 reports the exact optimum for the reference mean objective. P10 and alternative operators require separate optimization.
\end{minipage}\end{figure}

The absolute gain is small because the added capacity is small: 6.25 units are 0.13\% of the baseline capacity of the 4,748 pharmacies. For the same 6.25 units, the exact optimum yields 5.2 times the gain credited by the scalar model (0.00122), and the better reach-based proposal 2.3 times. Under the exact optimum, the gain is concentrated near the 25 selected pharmacies: 1.17 million daytime residents within their reach gain 0.063 Retention units on average (Supplementary Section S7.6).

\setcounter{subsection}{3}
\subsection{Operators, scale and shortlist size}

The operators agree on citywide ranks while selecting different local priority sets. Median pairwise Spearman correlation is about 0.81 (Fig. 10a), whereas Jaccard overlap is about 0.56 for bottom-decile sets (Table 6) and 0.59 for 5\% tail sets (Fig. 10b). Ward aggregation raises rank correlations to about 0.97. Yet ward-level membership disagreement exceeds grid-level disagreement in over half the cases (Supplementary Fig. S11). Higher aggregate agreement therefore leaves unresolved which individual cells need attention. In the daytime road-disruption example, 232 cells have contested membership (Fig. 11). Across the shared cases, bottom-decile memberships selected only by the reference operator concentrate near the cutoff; none lies deep in the tail (Supplementary Fig. S10).

The operators also agree on the citywide late-week decline; their differences are an order of magnitude smaller than the inventory substitution effect (Supplementary Table S8C).

Shortlist size changes operator sensitivity. Lists of 25 cells agree completely across operators under the common tie rule (Table 6). Larger lists agree less, placing more residents in contested members. Many cells tie at the smallest list's cutoff, so its apparent unanimity depends on the origin-identifier tie rule. A secondary spatial analysis finds no local cluster meeting the Benjamini–Hochberg criterion (minimum adjusted p-value 0.074; Supplementary Section S10).

\begin{figure}[!tp]\begin{minipage}{\linewidth}\centering
\includegraphics[width=170mm,height=180mm,keepaspectratio]{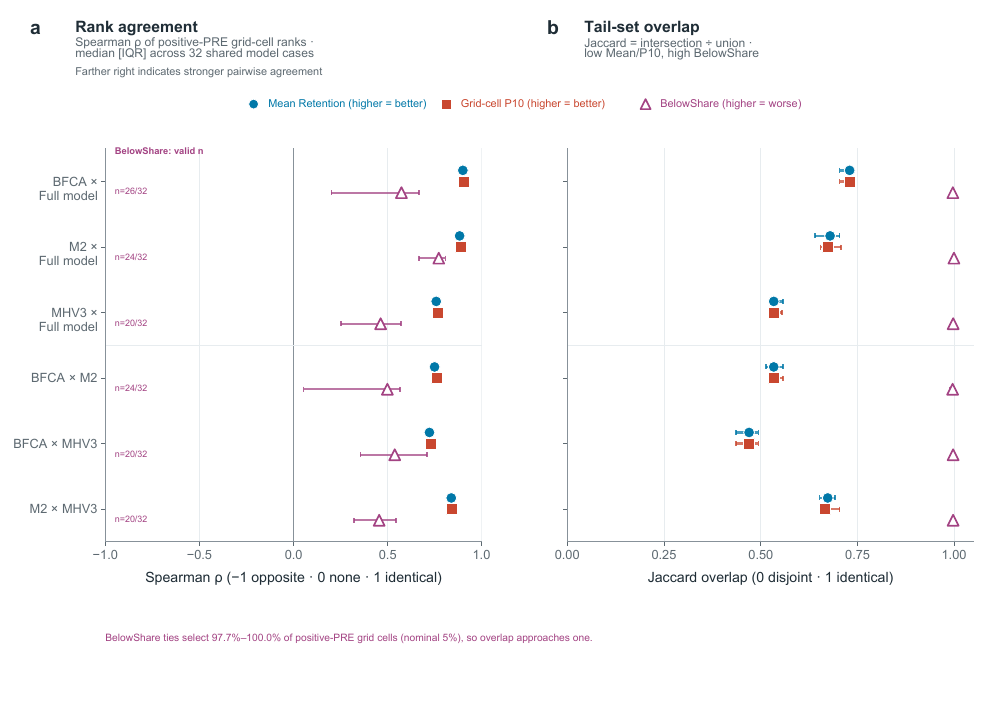}
\par\smallskip\raggedright\small Figure 10. Operator agreement across the shared cases. Panel (a) reports pairwise Spearman rank correlations; panel (b) reports Jaccard overlap of 5\% tail sets. The main-text comparison uses mean Retention and grid-cell P10. Triangles show BelowShare, the share of draws in which a cell's Retention is below 1, defined in Supplementary Section S4.6. Median values and interquartile ranges (IQR) summarize the case comparisons. M2 and MHV3 abbreviate M2SFCA and MHV3SFCA.
\end{minipage}\end{figure}

\begin{figure}[!tp]\begin{minipage}{\linewidth}\centering
\includegraphics[width=170mm,height=180mm,keepaspectratio]{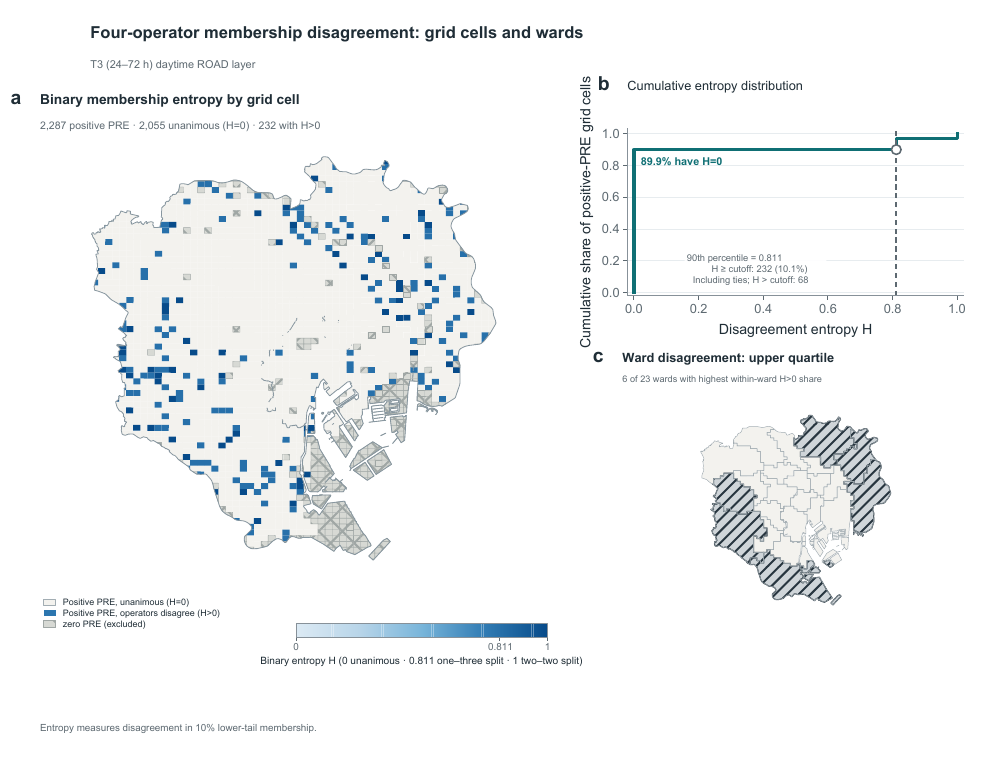}
\par\smallskip\raggedright\small Figure 11. Where operators disagree in the daytime road-disruption example. (a) Four-operator membership disagreement in the bottom-decile set, measured by binary entropy. Zero means unanimity; higher values indicate a split decision. (b) Cumulative distribution of entropy. (c) The six wards with the highest share of contested cells. Here H denotes membership entropy, distinct from the depletion horizon in Section 4.1. Table 6 reports the contested-cell count.
\end{minipage}\end{figure}

\setcounter{subsection}{4}
\subsection{Robustness: specifications, road closure, study boundary and resampling}

The reported directions hold under the tested changes to the specification, the road network, the study boundary and the sampled inputs, except when pharmacy loss shortens the depletion horizon. Among the eight specification families, changing pharmacy availability moves Retention across the widest part of the map: the median share of cells changing by at least 0.05 is 98.3\% (Fig. 12). Inventory-policy alternatives affect no cell in the first two phases and nearly every cell in the final phase, and median coverage for the soft road-cost alternatives is 2.6\% (Supplementary Section S6).

A hard-closure test removes every class-A road edge crossing a segment with blockage exposure above 0.10, about one-third of network length (Supplementary Section S2.2). The focal T3 and T4 means are then 0.315 and 0.179 under the proportional rule and 0.399 and 0.436 under the threshold rule, so the reversal between rules persists; the lower tail declines under both. The closure leaves 1,120 cells with positive PRE without a pharmacy within modeled reach.

Pharmacy loss lowers Retention, and it changes the late-week directions only when it also shortens the depletion horizon (Supplementary Section S6.5). With the horizon unchanged, the reversal between rules persists at ward quotas from 0.75 down to 0.15, under hazard-concentrated loss and under the opening hours listed in the register. With the horizon shortened in proportion to the share of pharmacies available, the late-week mean declines under both rules at every quota. The threshold-rule rise ends once the horizon is about 8\% shorter than its sampled value, whatever the number of pharmacies available. A count of available pharmacies therefore indicates the level of Retention but not its late-week direction.

Local tail membership responds most to where pharmacies remain available: it is stable under alternative daytime allocations and least stable under hazard-concentrated pharmacy loss (Table 6). Extending the domain beyond the wards changes focal mean Retention by less than 0.0001 and reduces unreachable cells from 364 to 347, while final-phase bottom-decile membership changes for 26,000–56,000 residents (Supplementary Section S9).

Independent redraws of the sampled inputs, with pharmacy-level random streams fixed, change population-weighted mean Retention by at most 0.00026 and the population-weighted average of grid-cell P10 by at most 0.0028. Every tested late-week path keeps its mean and P10 directions. At most three members enter and three leave a fixed-size list, with minimum Jaccard overlap 0.923 (Table 6; Supplementary Section S11.1).

\begin{figure}[!tp]\begin{minipage}{\linewidth}\centering
\includegraphics[width=170mm,height=180mm,keepaspectratio]{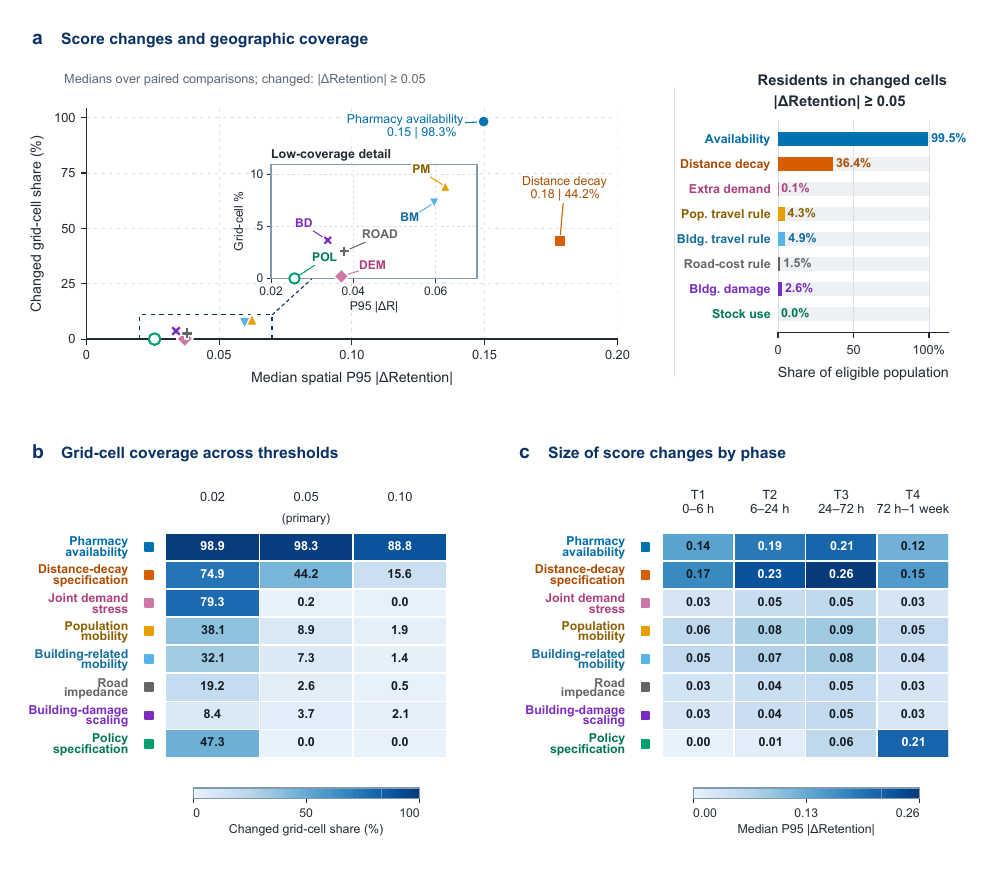}
\par\smallskip\raggedright\small Figure 12. Magnitude and reach across paired specification comparisons. (a) Median spatial 95th percentile (P95) of absolute Retention change against the share of cells changing by at least 0.05, with population coverage. (b) Coverage at the main threshold and at 0.02 and 0.10. (c) Median magnitude by phase. Family labels: BD, building-damage denominator; BM, building-based mobility; DEM, additional demand; PM, population-based mobility; POL, stock-use policy; ROAD, road costs. R abbreviates Retention. Families are different perturbations, not equal-cost interventions.
\end{minipage}\end{figure}

\begingroup
\scriptsize
\setlength{\tabcolsep}{2pt}
\renewcommand{\arraystretch}{1.12}
\begin{longtable}{>{\raggedright\arraybackslash}p{\dimexpr0.96\linewidth/5-2\tabcolsep\relax}>{\raggedright\arraybackslash}p{\dimexpr0.96\linewidth/5-2\tabcolsep\relax}>{\raggedright\arraybackslash}p{\dimexpr0.96\linewidth/5-2\tabcolsep\relax}>{\raggedright\arraybackslash}p{\dimexpr0.96\linewidth/5-2\tabcolsep\relax}>{\raggedright\arraybackslash}p{\dimexpr0.96\linewidth/5-2\tabcolsep\relax}}
\multicolumn{5}{@{}p{\linewidth}@{}}{\normalsize Table 6. Stability of citywide ranks and local shortlists under five sources of variation (32 shared cases unless stated).}\\[6pt]
\toprule
Source of variation & What is compared & Citywide agreement & Shortlist agreement & Residents in changed members \\
\midrule
\endfirsthead
\toprule
Source of variation & What is compared & Citywide agreement & Shortlist agreement & Residents in changed members \\
\midrule
\endhead
Operator & Reference operator, BFCA, M2SFCA, MHV3SFCA on grid cells & Median pairwise Spearman 0.805 (mean), 0.811 (P10) & Bottom-10\% Jaccard 0.561 / 0.563; 232 of 2,287 cells contested in the T3 road example & — \\
Aggregation scale & The same operators on ward totals & Spearman 0.967 / 0.968 & Ward-level membership disagreement exceeds grid-level in 59\% (mean) / 53\% (P10) of cases & — \\
List size (Supplementary Table S7C) & Operator pairs, fixed-size lists & — & Jaccard 1.000 at $K = 25$ (1–38 cells tied at the cutoff, median 31), 0.739 at $K = 50$, 0.613 at $K = 100$ & 0; 32,842; 125,731 \\
Coordinate redraw & Original set versus three ordinary LHS sets, 660 cases, pharmacy streams fixed & Maximum change 0.00026 Retention units (mean), 0.0028 (P10); all 486 T3–T4 directions kept & 2,052 of 2,304 fixed-size lists unchanged, ≤3 in / 3 out, min. Jaccard 0.923; 72/72 pharmacy sets and 54/54 placement preferences unchanged; reconfiguration replacements 76 → 73/76/74 & — \\
Input replacement (Section 5.5; Supplementary Table S7C) & Daytime allocation; capacity heterogeneity; concentrated versus random loss & — & Bottom-decile Jaccard 0.957–1.000; 0.722–0.847; 0.502 (mean) / 0.380 (P10) & — \\
\bottomrule
\end{longtable}
\endgroup

Note. Full redraw results, including additional outcomes and inventory-extension directions, are in Supplementary Table S5A.

\begingroup
\scriptsize
\setlength{\tabcolsep}{2pt}
\renewcommand{\arraystretch}{1.12}
\begin{longtable}{>{\raggedright\arraybackslash}p{\dimexpr0.96\linewidth/5-2\tabcolsep\relax}>{\raggedright\arraybackslash}p{\dimexpr0.96\linewidth/5-2\tabcolsep\relax}>{\raggedright\arraybackslash}p{\dimexpr0.96\linewidth/5-2\tabcolsep\relax}>{\raggedright\arraybackslash}p{\dimexpr0.96\linewidth/5-2\tabcolsep\relax}>{\raggedright\arraybackslash}p{\dimexpr0.96\linewidth/5-2\tabcolsep\relax}}
\multicolumn{5}{@{}p{\linewidth}@{}}{\normalsize Table 7. Sensitivity to the five sampled inputs: squared standardized rank regression coefficient ($\mathrm{SRRC}^2$), with Kolmogorov–Smirnov (KS) distance in parentheses.}\\[6pt]
\toprule
Input or diagnostic & T1 & T2 & T3 & T4 \\
\midrule
\endfirsthead
\toprule
Input or diagnostic & T1 & T2 & T3 & T4 \\
\midrule
\endhead
Depletion horizon & 0.000 (0.042) & 0.005 (0.036) & 0.245 (0.346) & 0.968 (0.794) \\
Staffing & 0.706 (0.587) & 0.637 (0.544) & 0.507 (0.445) & 0.017 (0.029) \\
Operation & 0.296 (0.462) & 0.359 (0.474) & 0.228 (0.332) & 0.008 (0.071) \\
Minor-injury demand & 0.000 (0.059) & 0.000 (0.063) & 0.000 (0.066) & 0.000 (0.039) \\
Affected-only demand & 0.001 (0.055) & 0.002 (0.065) & 0.003 (0.066) & 0.001 (0.038) \\
Rank-regression $R^2$ & 0.982 & 0.977 & 0.947 & 0.986 \\
Unexplained rank-response variation, $1 - R^2$ & 0.018 & 0.023 & 0.053 & 0.014 \\
\bottomrule
\end{longtable}
\endgroup

Note. Conditional on the prescribed input ranges. $\mathrm{SRRC}^2$ describes rank-regression sensitivity; coefficients are non-additive. $R^2$ is the rank-regression coefficient of determination. Residual variation includes nonlinearity, interactions and pharmacy-level randomness. KS compares the coordinate distribution among lower-tail cell–draw pairs with its uniform marginal.

\setcounter{section}{5}
\section{Discussion}

\setcounter{subsection}{0}
\subsection{What the three findings mean for recovery judgments}

Operating recovery and sustained service follow different paths when pharmacies consume finite stocks. The Noto supply interruptions illustrate this distinction \citep{tai2024disasterpharmacists}, which also appears in functional-accessibility accounts of service availability \citep{enderami2024postdisaster}. Here proportional and threshold rules produce opposite mean trends under the same stock inputs. Within one stock-balance process, the mean improves at phase entry but declines when averaged over the phase. Recovery statements in Retention therefore need both a service rule and a measurement timing. Temporal variation matters for routine services \citep{park2026customized} and for post-earthquake access. Routine pharmacy studies show how opening hours reshape access \citep{clark2025pharmacy,viana2025identifyingpharmacy}. This study adds stock as a state variable in effective capacity for city-scale post-earthquake access to pharmacy-based OTC medicines. That contribution extends the Tokyo evidence on emergency-vehicle accessibility \citep{kishimoto2019accessibility} and routine pharmacy proximity \citep{yamazaki2026healthcare}.

The reduction tests identify the information needed for each task. Shared inventory trajectories make the citywide maps amenable to scalar approximation. Reconfiguration changes where capacity remains, requiring the spatial links that a fixed map lacks. This complements work on facility configuration in catchment models \citep{delamater2013spatial}. The operator's linearity in supply explains why the mean-input spatial calculation performs well, while sampled demand prevents exact equivalence. A practical workflow can use scalar profiles for initial screening, spatial mean inputs to compare support locations, and full sampling to estimate lower-tail outcomes.

Broad rank agreement can coexist with different local priority lists. Demand allocation and provider choice \citep{paez2019demand,jorg2023mhv3sfca} move cells across cutoffs with little change in citywide rank correlation. This connects the local measurement differences reported by Apparicio et al. \citeyearpar{apparicio2008comparing} with the scale-dependent parameter sensitivity found by Chen and Jia \citeyearpar{chen2019comparative}. Here the disagreement concentrates near cutoffs and persists under aggregation and independent coordinate redraws. The spatial evidence is strongest in the response to pharmacy reconfiguration and in contested memberships; adjusted local-cluster tests yielded no localized disaster hotspots.

\setcounter{subsection}{1}
\subsection{Use in Tokyo and transfer to other cities}

Tokyo's disaster pharmaceutical coordinators collect service information and coordinate medicine supply and pharmacist activities \citep{tokyocoordinator2025}. In March 2026, Tokyo introduced a mobile pharmacy jointly with the Tokyo Pharmaceutical Association \citep{tokyo2026mobilepharmacy}. The vehicle is stationed at the Tachikawa Regional Disaster Prevention Center and supports dispensing and medicine supply at disaster medical relief stations. The placement comparison shows the scale at which such support acts. Optimally placing 6.25 facility-equivalent units raises focal mean Retention by 0.00632, about 3\% of the proportional-rule late-week decline. In the same convention and profile, restoring the T3 inventory state alone raises the T4 mean by 0.317 (Fig. 4d). Capacity additions at the scale of a few units therefore act locally; the citywide trend is governed by stock. At fixed travel weights, increments at existing pharmacies cannot reach the 364 cells without modeled pre-disaster access or the 1,120 positive-PRE cells that lose reach under hard closure; mobile service is a relevant option for these cells. Within the prescribed input ranges, the final-phase $\mathrm{SRRC}^2$ is 0.97 for the depletion horizon and 0.02 for staffing (Table 7). Status reports limited to opening and staffing therefore omit the input that dominates late-week sensitivity. A stock-status field would reveal service losses that operational recovery conceals.

Under the threshold rule, the citywide mean rises while the population-weighted average of grid-cell P10 falls, because P10 reflects adverse conditions, chiefly short depletion horizons. A mean-only verdict can therefore signal recovery while adverse-condition access continues to worsen.

A preparedness review has four steps. First, map pre-disaster accessibility and the 364 cells with no modeled pharmacy access. Second, compare both service rules and both measurement timings alongside their lower tails. Third, form candidate pools from locations recurring across operators and flag memberships near cutoffs. Fourth, rank existing pharmacies by their exact marginal gain $g_j$ for equal-capacity additions under the mean objective, then evaluate lower-tail outcomes and equity separately. Retention should be read alongside absolute accessibility and population coverage: a large proportional loss and a low absolute service level describe different needs.

Three parts of the analysis transfer to other cities. First, the inputs are of kinds that many cities hold: a provider register, gridded population, a walking network, a hazard scenario with damage estimates, and stated ranges for stock, staffing and demand. Second, three results follow from the structure of the calculation, not from Tokyo's geography. A stock fraction shared across providers factors out of the calculation, so the citywide verdict depends on citywide inputs: the stock-to-service rule, the depletion horizon relative to the assessment period and the pace of staffing recovery. A fixed map rescaled by citywide totals cannot register where capacity moves, whereas the gain from added capacity at each provider has an exact expression. Third, the reporting practice applies generally: state the stock-to-service rule and measurement timing, and report lower-tail outcomes with the mean. The numerical results are specific to Tokyo, which has one pharmacy per about 2,000 nighttime residents. Agreement between methods, stability of priority lists and gains from placement should be re-estimated for cities with sparser networks, car-based access or slower replenishment.

\setcounter{subsection}{2}
\subsection{Limitations}

The model has not been validated against observed post-disaster outcomes. No records of pharmacy opening, stock, sales or deliveries after an earthquake were available to this study. The results therefore show which modeling choices change a recovery judgment; they do not predict the level of access after a real earthquake. The 0.5 level used to report low Retention is likewise declared, not validated as an adequacy threshold: the spare capacity of pharmacies in normal times is unobserved.

The model evaluates OTC access, one component of pharmacy service continuity. The same network dispenses prescriptions, including disaster prescriptions \citep{koike2026pharmaceuticalneeds}, so its continuity also bears on treatment maintenance, which this model does not evaluate. The provider register covers licensed pharmacies, one part of the OTC retail network. Store-sales outlets without a pharmacy license are absent. Their omission can make neighborhoods appear to have fewer OTC options and can change priority membership. Because Retention divides post-disaster access by its own baseline, adding these outlets could affect both numerator and denominator; the direction of bias in the ratio is unknown.

Demand stays at residential or daytime origins throughout the week. Evacuation could reduce demand in damaged residential cells and concentrate it near designated shelters, altering travel requirements and competition for nearby pharmacy capacity. Shelter-based medical teams and medicine distribution would also add service channels outside the modeled pharmacy network. The present maps therefore cannot identify shelter-specific shortages or assess the adequacy of shelter medicine provision.

Preparedness ranges govern the results. The reference cases keep every registered pharmacy available at reduced capacity; pharmacies are unavailable only in the quota comparisons and in one check based on the opening hours listed for normal operation. In these comparisons no pharmacy reopens during the week. How much demand moves to the available pharmacies is not modeled; the quota comparisons take none and all as its bounds. The 3–7-day depletion horizons place depletion within the study week, and the reference shares one stock multiplier across pharmacies in each draw. Heterogeneous-stock extensions relax that assumption, but product matching, post-stockout rerouting and calibrated delivery networks remain unmodeled. The population and damage inputs are from 2015 and 2022, respectively, and the map that orders the daytime additions is from 2026. Combining them with the 2025 pharmacy register adds temporal mismatch. Between the 2015 and 2020 censuses the nighttime population of the 23 wards grew by 5.0\% and the daytime population by 3.6\% \citep{tokyo2024daytimecensus}. Uniform growth leaves Retention unchanged, because pre- and post-disaster accessibility use the same population; uneven growth would shift local values and ranks. Active fires and debris in burnt areas do not close routes in the network, which can overstate early walking access within fire-spread zones.

Four-operator comparisons cover 32 shared cases. Inventory-rule and reduction tests use the reference operator; their quantitative effects await evaluation with the comparators. The exact placement optimum applies to the fixed-demand, fixed-weight mean objective and the stated equal-increment constraint. Travel feasibility, delivery costs and equity objectives would change that allocation problem. External-domain tests use transferred disaster states.

\setcounter{subsection}{3}
\subsection{Future work}

Two extensions follow directly. First, records of pharmacy operation, stock, deliveries and sales after an event would allow validation against reported status, for example through pharmacy status-reporting systems \citep{mitoya2023creationfirst}, and would support calibrated marginal distributions and joint dependencies. Observed travel and demand patterns would then help select the service mapping and turn horizon–addition boundaries into testable stocking conditions. Mobile capacity is a further placement case: a candidate relief-station site can be ranked with the same marginal-gain expression once its travel weights are computed, whereas deployment depends on vehicle passability \citep{kishimoto2019accessibility}, which this walking model does not represent. Second, a shelter scenario could relocate a stated share of demand from heavily damaged cells to designated evacuation sites. Adding store-sales outlets when a geocoded register becomes available would test the effect of broader OTC coverage. Together, these two scope extensions would show how the current local priorities depend on population location and provider coverage.

\setcounter{section}{6}
\section{Conclusion}

This study added pharmacy stock as a state variable to a city-scale accessibility model and asked which modeling choices change judgments about first-week medicine access after a Tokyo earthquake. Three conclusions follow. First, whether access appears to recover depends on two modeling assumptions: how remaining stock translates into service, and whether recovery is measured at the start of a period or averaged over it. Mean and low-end (P10) outcomes can also move in opposite directions under the same assumptions. Second, the level of model detail should match the planning task. Citywide totals suffice to approximate citywide maps. Evaluating pharmacy closures or the placement of added capacity requires spatially resolved inputs. Estimating low-end outcomes requires sampled inputs. Third, alternative catchment methods agree on citywide rankings but select different local priority areas. Local shortlists should therefore be treated as method-dependent candidates for verification rather than definitive targets. For Tokyo's preparedness, these results suggest that status reporting should include stock levels alongside opening and staffing. Assessments should also report low-end outcomes, not only means. The same tests can be run in other cities and for other services that depend on stock, such as food or fuel distribution after disasters. Calibrating the model with observed stock, delivery and sales records, and extending it to store-sales outlets and shelter demand, are the next steps toward operational use.

\section*{Acknowledgments}

This research was the result of the joint research with CSIS, the University of Tokyo (No. 1361) and used the following data: Pseudo People Flow: Personal Attribute Dataset (Tokyo) provided by CSIS People Flow Project Office. We are grateful to participants at the CSIS DAYS 2026 (Kashiwa, Japan, 13–14 November 2026) for their helpful comments.

\section*{Funding}

This work was supported by JSPS KAKENHI Grant Numbers JP20K14898 and JP26K22096.

\section*{Declaration of competing interest}

The authors declare that they have no known competing financial interests or personal relationships that could have appeared to influence the work reported in this paper.

\section*{CRediT authorship contribution statement}

Juanning Liang: Conceptualization, Data curation, Formal analysis, Investigation, Methodology, Software, Validation, Visualization, Writing – original draft, Writing – review \& editing. Hiroki Baba: Conceptualization, Funding acquisition, Supervision, Writing – review \& editing.

\section*{Data and code availability}

The official earthquake damage estimates (Tokyo Metropolitan Disaster Prevention Council, 2022), the MHLW Medical Information Net pharmacy register, the 2015 census grid statistics (Statistics Bureau of Japan, 2015) and the 2021 building-conditions survey (Tokyo Metropolitan Government Bureau of Urban Development, 2021) are publicly available from the cited sources. The synthetic people-flow data were obtained under a joint research agreement with the Center for Spatial Information Science (CSIS), The University of Tokyo. Access was provided through the Joint Research Assist System (JoRAS; project no. 1361). The agreement does not permit redistribution of these data; qualified researchers can request them from CSIS. Result tables at grid-cell and pharmacy level are derived from these data and are not released; aggregate results are reported in the Supplement. The pharmacy-provenance crosswalk and the comparison registers summarized in the Supplement will be deposited in a public repository upon acceptance. The deposit will also include the NumPy/SciPy analysis code for the reference model and comparators, with a synthetic test case that runs without restricted inputs. A digital object identifier will be added after deposition.

\section*{\raggedright Declaration of generative AI and AI-assisted technologies in the manuscript preparation process}

During the preparation of this work the authors used ChatGPT (OpenAI) in order to polish the language, including grammar and wording, and to assist with implementing and completing analysis code. After using this tool, the authors reviewed and edited the content as needed and take full responsibility for the content of the published article.
\bibliographystyle{journalplainnat}\bibliography{references}
\end{document}